\RequirePackage{lineno} 
\documentclass[preprint,aps,prd,nofootinbib,superscriptaddress,showpacs, showkeys]{revtex4}

\usepackage{graphicx} 
\usepackage{subfigure}
\usepackage{amsmath}
\usepackage{epstopdf}

\newcommand{\Eslash}{\mbox{$\rm E \kern-0.6em\slash$                }}
\newcommand{\pslash}{\mbox{$\rm p \kern-0.6em\slash$                }}
\newcommand{\etmiss}{\mbox{$\rm \Eslash_{T}\!$                        }}
\newcommand{\ptmiss}{\mbox{$\rm \Eslash_{T}^{trk}\!$                        }}

\begin{document}
\title{Searching for Dark Matter at Hadron Colliders}
\author{Andrew Askew}
\affiliation{Department of Physics, Florida State University, Tallahassee, FL, 
USA}
\author{Sushil Chauhan}
\affiliation{Department of Physics, University of California Davis, Davis CA, 
USA}
\author{Bj\"orn Penning}
\affiliation{Fermilab National Laboratory, Batavia IL, USA and The University of 
Chicago, Chicago, IL, USA}
\author{William Shepherd}
\affiliation{Santa Cruz Institute for Particle Physics and Department of 
Physics, University of California Santa Cruz, Santa Cruz, CA, USA}
\author{Mani Tripathi}
\affiliation{Department of Physics, University of California Davis, Davis CA, 
USA}

\date{\today}

\begin{abstract}
Theoretical and experimental techniques employed in dedicated searches for dark 
matter at hadron colliders are reviewed.  Bounds from the 7 and 8 TeV 
proton-proton collisions at the LHC on dark matter interactions have been 
collected and the results interpreted. We review the current status of the
Effective Field Theory picture of dark matter interactions with the Standard Model.
Currently, LHC experiments have stronger bounds on operators leading to
spin-dependent scattering than direct detection experiments, while direct detection
probes are more constraining for spin-independent scattering for WIMP masses above a few GeV.
\end{abstract}

\maketitle

\section{Introduction}
\label{sec:intro}

With the discovery of the Higgs boson at the LHC \cite{Aad:2012tfa, 
Chatrchyan:2012ufa} the problem of electroweak symmetry breaking has been 
solved, albeit superficially. Significant theoretical reasons remain for 
the involvement of additional physics in the electroweak sector, chief among 
them the sensitivity of the Higgs mass to any new higher scale introduced to our 
theory. However, the experimental demand for new particles to regulate weak boson 
scattering, pending confirmation of the Higgs properties, appears to have been 
resolved. While there remains an expectation that there should be new particles at the TeV scale, 
the Standard Model (SM) continues to 
describe all the measurements made at colliders throughout history surprisingly well.

The next great experimental requirement for new physics is due to an entirely 
different sort of measurement. Particle physics has historically been driven by 
looking at interactions on ever smaller length scales, where new interactions 
were required in order to understand the behavior observed at slightly longer 
distances. The great new arrival on the scene for particle physics is the 
overwhelming evidence for the existence of dark 
matter (DM). Various evidences and inferences which lead to the widely accepted 
existence of this fundamentally new component of the universe have been laid out 
in great detail in \cite{Bergstrom:2012fi}.

The only constraints that can be placed on dark matter from cosmological 
concerns are that it have the correct cosmological energy density, that it be 
massive so that it can act (at least for gravitational purposes) as pressureless 
matter, and that it not interact so strongly as to either disturb the 
well-understood cosmic microwave background 
\cite{Ade:2013zuv,Madhavacheril:2013cna} or to fail to collapse sufficiently to 
explain the observed large-scale structure of the universe 
\cite{Madhavacheril:2013cna}. While these requirements are fairly 
straightforward to state, much effort has been invested in understanding their 
precise implications for the particle physics of dark matter.

Many different models and mechanisms for the properties and production of dark 
matter exist in the early universe, with the masses predicted for dark matter 
ranging from the $\mu$eV scale to beyond the Planck scale. Each of these 
different models has distinct motivations and assumptions for the physics of the 
early universe.

For some time the most popular models of DM have been based on the so-called 
Weakly Interacting Massive Particle (WIMP) Miracle \cite{Feng:2008ya}. The basic 
observation is that if DM is in thermal equilibrium in the early universe and 
leaves equilibrium only when its interactions become rare compared to the Hubble 
rate then the mass and annihilation cross section of DM are enough to determine 
the remaining `thermal relic' abundance of DM in the universe. The miracle is 
that, by simply assuming that DM is some stable particle which is involved in 
electroweak symmetry breaking, one finds the correct order of magnitude for the 
DM abundance. Since neutral, stable particles are fairly common in models 
introduced to explain other theoretical problems, notably the gauge hierarchy 
problem, this has been considered to be a very promising mechanism for DM 
production.

Guided by this consensus that WIMP DM candidates are particularly 
well-motivated, a robust experimental effort is underway to either discover DM, 
or constrain candidate theories. It is widely agreed that, in order to develop a 
reasonably complete understanding of the physics of DM, multiple different 
detection and discovery techniques are required. The two main avenues which have 
concentrated on the DM specifically are generally termed direct and indirect 
detection, depending on whether the search is for WIMPs themselves or their 
annihilation products.

Direct detection involves searching for WIMPs scattering off of nuclei of 
ordinary matter. These scatterings impart small amounts of kinetic energy 
(typically $<$ 100 keV) to the recoiling nuclei, thus making it necessary to build 
detectors capable of low energy detection thresholds.  Such experiments are 
deployed in laboratories with a large overburden of earth, and are instrumented 
within elaborate shields against cosmic ray interactions in order to remain effectively background-free. For 
direct searches, the energy threshold, target mass and exposure time become the 
limiting factors governing the sensitivity of the experiments. This scattering 
rate can be related by the particle physics model of DM to the annihilation 
cross section, which in turn must have the appropriate behavior to explain the 
DM energy density of the universe.

Indirect detection, by contrast, involves searches for the products of the very 
annihilation processes that are responsible for establishing the DM relic 
density.  These processes should still be going on today, especially in regions 
where the local DM densities are much higher than they are in our galactic 
neighborhood. The unfortunate disadvantage associated with indirect detection is 
the background induced by `ordinary' astrophysics. For example, If one were to 
consider only the signal strength from DM annihilations, by far the most 
promising target for indirect detection would be the galactic center, but there 
is a relatively unconstrained astrophysical background to any signals that could 
be produced there. Searches in the Galactic Center have indeed found new contributions which are consistent with a DM annihilation origin \cite{Goodenough:2009gk,Hooper:2010mq,Hooper:2011ti,Abazajian:2012pn,Su:2012ft,Daylan:2014rsa}, but the signals are also compatible with astrophysical backgrounds and/or statistical fluctuations. Dwarf spheroidal galaxies, which are satellites of the Milky 
Way, are very rich in DM content and provide ideal sources for indirect 
searches, however, all attempts at measuring an excess of gamma rays in those systems has yielded 
null results \cite{Ackermann:2013yva}.

Complementary to these two approaches, collider searches \cite{Mitsou:2013rwa} have historically 
focused on the other aspects of the particle theories that happened to contain 
DM candidates, searching for their strongly-coupled new particles and using the 
DM candidate as a convenient tag for new physics events in the form of missing 
energy. Recently, a more direct focus on DM at colliders was proposed 
\cite{Goodman:2010yf} and rapidly adopted \cite{Goodman:2010qn, Fox:2011fx, 
Rajaraman:2011wf, Fox:2011pm, Friedland:2011za, Cheung:2012gi, Aaltonen:2012jb, 
Fox:2012ee, Bai:2012xg, Chae:2012bq, Fox:2012ru, Carpenter:2012rg, Zhou:2013fla, 
Cornell:2013rza, Cotta:2013jna, Buckley:2013jwa, Essig:2013vha, Agrawal:2013hya, 
Crivellin:2014qxa, Busoni:2014sya, Mao:2014rga, Davidson:2014eia, Lopez:2014qja, 
Artoni:2013zba} by the collider community. The goal is to relate the pair 
production rate of DM at colliders to the annihilation and scattering rates at 
more traditional DM-oriented experiments, while making as few assumptions about 
other, possibly related new physics as possible. This approach is made possible 
in large part by applying effective field theory techniques to the problem of DM 
interactions.

Collider searches for DM have their own advantages and weaknesses. Both direct 
and indirect detection signals lose strength as the DM candidate becomes lighter 
due to the smaller amount of energy available in each interaction. In contrast, 
colliders are able to copiously produce light particles, and hence do not suffer 
from threshold effects in their search for low-mass DM. However, they suffer 
from parton distribution function suppression for high DM masses (above hundreds of GeV), where the 
other two search techniques are more robust, even though the rates are lowered 
due to the reduction in DM number density. The greater uncertainty associated 
with DM searches at colliders, however, is whether or not a putative signal is 
actually caused by the true DM in the universe or an imposter particle that is 
stable only on collider timescales, but not cosmological ones. This is not at 
all a concern in other techniques, as they are looking for the currently extant 
DM particles, rather than producing a new pair of particles.

In this article, we begin by reviewing the effective field theory and other 
theoretical tools which have been employed in designing and interpreting 
collider searches for dark matter (section \ref{sec:effth}). We proceed to 
outline the general experimental approaches employed in searching for direct 
dark matter production (section \ref{sec:exp}), followed by a discussion of the 
current leading results derived from those searches at the LHC and elsewhere 
(section \ref{sec:results}). 


\section{Model Independent Dark Matter Interactions}
\label{sec:effth}

When trying to understand DM interactions with ordinary matter, it can be useful 
to `de-focus' one's theoretical eye a bit, and consider generic interactions 
rather than specific models of DM. This, hopefully, helps us understand that 
which is truly generic to DM physics as opposed to the specific predictions of 
one particular model. A fully general approach, assuming nothing about DM except 
that it exists, has some mass, and is (at least phenomenologically) stable, is 
achieved by working in an effective field theory framework and writing down 
interactions between the DM and SM fields of interest. An exhaustive approach to 
interactions of DM with hadronic matter was undertaken in \cite{Goodman:2010ku}. 
Here we will choose a smaller but representative set of possible interactions to 
consider. We will consider scalar and fermionic DM particles, assuming always 
that the DM is completely uncharged under the SM gauge group, and look only at 
those interactions that can induce appreciable signals at tree level in direct 
detection experiments. Note that, while in principle it is possible for scattering off
of leptons to contribute to at least some direct detection searches, these contributions
are strongly suppressed compared to nuclear scatterings by the kinematics involved.
To avoid possible confusion, we will refer to fermionic 
DM candidates uniformly as $\chi$ and scalar DM candidates as $\phi$.

In the spirit of healthy scientific agnosticism, it must be acknowledged that 
even these most basic assumptions about the physics of DM are subject to 
argumentation. Many models exist where DM is charged under the SM weak 
interactions. In fact, these models are the original WIMP models, where the DM 
explicitly interacted with the weak force. Other models consider DM that is 
milicharged under electromagnetism \cite{Goldberg:1986nk}. In general, the 
model-independent technique strives to understand the physics of all models as 
much as is possible, however, there will always be exceptions to its 
applicability.

The representative interactions which we will consider are listed in table 
\ref{tab:ops}. Note that they have names which originated in 
\cite{Goodman:2010ku} and can be considered purely historical. Each operator is 
preceded by an assumed Wilson coefficient, and we have chosen those which are 
most standard in the literature.

The scalar-type operators D1 and C1 are of
a higher dimensionality than they na\"{\i}vely appear, with an additional suppression of $m_q/\Lambda$.
This factor is chosen for two reasons. First, the scalar-type operators violate $SU(2)_L$, and thus
technically must also couple to the Higgs boson to be gauge invariant. Replacing the Higgs by its vacuum expectation value gives
a form similar to the one we have chosen. The choice to scale the operators by each quark's mass rather than
by a uniform factor of $v/\Lambda$ is motivated by the conjecture of Minimal Flavor Violation (MFV) \cite{Buras:2000dm}, which protects
new models from being strongly constrained by flavor physics observables by insisting that all flavor violation be
proportional to the SM Yukawa matrices. It is worth noting that the normalization of D9 
actually isn't very well-motivated from a theoretical point of view. In 
principle, the same suppression by $m_q$ that is present for the scalar-type 
operators, D1 and C1, should also be present there, since the operator violates 
$SU(2)_L$ in the same way. This 
normalization for the tensor operator is standard, however, because this is the 
normalization which is most naturally probed in direct detection experiments. We 
have listed only the operators whose contributions to direct detection 
scattering are not suppressed by the small dark matter velocity. It is worth noting, however,
that other operators are possible and have bounds similar to those derived for these operators
from collider searches, while they are effectively unbounded by direct detection due to the
suppression of the scattering cross section.

\begin{table}
\subtable[Operators for Dirac fermion DM]{\begin{tabular}{|l|c|c|c|}
\hline
Name&Operator&Dimension&SI/SD\\
\hline
D1&$\frac{m_q}{\Lambda^3}\bar\chi\chi\bar qq$&7&SI\\
D5&$\frac{1}{\Lambda^2}\bar\chi\gamma^\mu\chi\bar q\gamma_\mu q$&6&SI\\
D8&$\frac{1}{\Lambda^2}\bar\chi\gamma^\mu\gamma^5\chi\bar 
q\gamma_\mu\gamma^5q$&6&SD\\
D9&$\frac{1}{\Lambda^2}\bar\chi\sigma^{\mu\nu}\chi\bar q\sigma_{\mu\nu}q$&6&SD\\
D11&$\frac{\alpha_s}{\Lambda^3}\bar\chi\chi G^{\mu\nu}G_{\mu\nu}$&7&SI\\
\hline
\end{tabular}}
\qquad
\subtable[Operators for Complex scalar DM]{\begin{tabular}{|l|c|c|c|}
\hline
Name&Operator&Dimension&SI/SD\\
\hline
C1&$\frac{m_q}{\Lambda^2}\phi^\dagger\phi\bar qq$&6&SI\\
C3&$\frac{1}{\Lambda^2}\phi^\dagger\overleftrightarrow{\partial}_\mu\phi\bar 
q\gamma^\mu q$&6&SI\\
C5&$\frac{\alpha_s}{\Lambda^2}\phi^\dagger\phi G^{\mu\nu}G_{\mu\nu}$&6&SI\\
\hline
\end{tabular}}
\caption{\label{tab:ops}Lowest-dimensional operators which couple singlet DM 
candidates to hadronic matter and give unsuppressed contributions to direct 
detection scattering of DM off of a nucleus. The fourth column indicates whether the primary direct 
detection signal due to that operator is spin-independent (SI) or spin-dependent 
(SD). As scalars have no spin, all the listed operators for scalar DM give rise 
only to SI direct detection signals.}
\end{table}

Operators of this type, in some combination, suffice to describe all the 
interactions of DM with hadronic matter, provided that the new particles 
involved in the interactions (of an extended theory) are much heavier than the 
scales at which we are probing the interactions between the two. This is 
manifestly the case in direct detection, where the probe energy is a tiny 
fraction of the DM mass due to the low velocity of DM in the galactic halo. 
However, it is much more suspect for a collider search, where the dark matter 
pair is generally produced relativistically and thus can have kinetic energy 
comparable to its mass. Nonetheless, these conditions can be met in many models, 
and these operators provide a good understanding of the physics of those 
scenarios.

One critical feature of these interactions which is often overlooked is their 
mixing. In particular, D1 and C1, which are very weakly constrained at the LHC 
due to the explicit quark mass suppression in the operator, can induce (at 
one-loop level) D11 and C5, respectively. The diagrams which cause this mixing 
are completely analogous with those responsible for the gluon fusion production 
of the Higgs boson \cite{PhysRevLett.40.692}. The latter operators 
are very strongly constrained by colliders, which highlights the importance of 
tracking these effects. This was originally pointed out in \cite{Haisch:2012kf}, 
wherein the effects of the full one-loop production are also calculated, outside 
of the heavy-quark limit in which one can truly treat the operators as mixing 
with no new form factors. They find that, while the bounds from the 
gluon-coupling operators cannot be directly transferred onto the scalar-type 
operators, they do not overstate the bounds by more than about a factor of 20 in 
cross-section (which is only a factor of 1.6 or 2.1 in the suppression scale of 
D11 or C5, respectively), still giving much stronger results than the scalar 
operator alone at tree-level.

Multiple approaches to understanding the range of applicability of these 
operators have been developed. One can require that the coupling between the 
particle needed to induce the interaction and the particles under consideration 
must be small enough to be perturbative as a first step, as is done in 
\cite{Goodman:2010ku}. More sophisticated approaches to understanding the range 
of applicability of these theories utilize the requirement of unitarity 
\cite{Shoemaker:2011vi}, which is generally a weaker but more robust constraint 
than the former, or have taken the approach of considering explicit models with 
relatively light interaction-mediating particles \cite{Busoni:2013lha, 
An:2013xka, DiFranzo:2013vra, Buchmueller:2013dya, Chang:2013oia}, a technique 
which has since been adopted by the CMS collaboration as well \cite{CMS-PAS-EXO-12-048}

The ``light mediator'' models are really examples of simplified models as recommended in \cite{Alves:2011wf}, where it is assumed that only one new particle (beyond 
the DM candidate) describes accurately all of the interactions of DM with 
ordinary matter. Generally, each of the operators in Table \ref{tab:ops} is 
replaced by a propagating particle, either scalar or vector, which has 
appropriate couplings to the DM candidate and to hadronic matter. The operator 
can again be inserted in the limit where the mediator mass becomes very large. 
As a simple example, the Lagrangian for a simplified model completion of 
operator D1 would be

\begin{equation}
\mathcal{L}\supset 
m^2_\Phi\Phi^\dagger\Phi+g_\chi\Phi\bar\chi\chi+g_q\frac{m_q}{m_\Phi}\Phi\bar qq 
+ {\rm h.c.},
\end{equation}

where $\Phi$ is the new mediator particle and the factor $\frac{m_q}{m_\Phi}$ is just a normalization
introduced into the quark coupling to maintain the MFV conjecture value for the 
interaction. It is straightforward 
to see that in the limit of large $m_\Phi$ this gives back the operator D1, with 
the assignment $\Lambda^3\equiv\frac{m_\Phi^3}{g_qg_\chi}$. One of the chief 
uncertainties in a model of this type is due to the width of the mediator 
particle. The standard assumption that is made is that the width scales linearly 
with the mass of the mediator, as one would expect on na\"{\i}ve dimensional 
grounds. However, it was shown in \cite{Goodman:2011jq} that the choice of 
behavior for the width has important implications for the strength and behavior 
of the bounds. We also note in passing that the narrow-width approximation is 
generically used for these particles, and in some cases has been extrapolated 
beyond its realm of applicability. In particular, conclusions have been reached 
that the bounds from a model with a very heavy mediator are actually weaker than 
those from the effective operator due to effects of the mediator width, but 
these are misleading, as the width of a far off-shell particle is not identical 
to the width of the same particle on-shell, and needs to be recomputed at the 
appropriate invariant mass.

Finally, it is important to note that even these simplified models can be overly 
simple from the point of view of DM physics. As explored in 
\cite{Profumo:2013hqa}, the dark matter's various interactions with SM particles 
are very commonly dictated by interactions with non-hadronic particles. In fact, 
it is often some mixture of different particles which is important, with the 
dominant species changing depending on the particular type of process we are 
considering. For example, a `vanilla' SUSY neutralino with the correct relic 
density generated by the well-tempering mechanism \cite{ArkaniHamed:2006mb} 
generically annihilates preferentially into weak vector bosons, but scatters in 
direct detection primarily through the Higgs boson. All of the approaches above, 
even if extended to consider interactions with either gauge or Higgs bosons, as 
done by \cite{Cotta:2012nj, Chen:2013gya}, have implicitly assumed that, for all 
the processes we are interested in, the DM interacts dominantly with the same SM 
field. Thus we emphasize again the importance of verifying the assumptions made 
in the model-independent results we quote below when trying to apply them to a 
given more-complete model of DM physics. Ultimately, when a complete model is 
under consideration, fully focused theoretical vision is best.


\section{Collider Searches for Dark Matter}
\label{sec:exp}

    Common to all DM searches is the signature of missing transverse momentum 
(often called missing transverse energy, \etmiss)~caused by the WIMPs escaping 
the detector. As described in Sec.~\ref{sec:effth} these events can 
be produced in association with Standard Model particles, most notably photons 
and jets (either from quarks or gluons) but also $W$, $Z$ or even Higgs bosons 
and heavy quarks ($b$- and top-quark). These particles produced in association with 
the WIMP pair will recoil against the invisible particles. The common signature 
is therefore a large value of \etmiss~and a back-to-back topology between 
\etmiss~and the Standard Model particle used for tagging.
In the ATLAS and CMS experiments analyses using jets, photons and $W$/$Z$ bosons 
have been performed with more final states yet to be analyzed. In the following 
we will briefly describe the existing analyses ordered according to their 
signature.  In all analyses only data which has passed the relevant quality 
criteria of ATLAS and CMS have been used in the searches reviewed.
 
  The direct detection experimental results from LUX~\cite{Akerib:2013tjd}, 
SIMPLE~\cite{Felizardo:2011uw}, SuperCDMS~\cite{Agnese:2014aze}, 
IceCube~\cite{Aartsen:2012kia} (assuming for concreteness a $W^+W^-$ dominant
annihilation channel), PICASSO~\cite{Archambault:2012pm}, and 
XENON-100~\cite{Aprile:2013doa}, which represent the best current bounds for 
WIMP-nucleon scattering, are also compared to get a complete overview of the 
current status of DM searches, including limits on WIMP-nucleon cross section 
and scale of these interactions ($\Lambda$).

\subsection{Mono-photon searches}
ATLAS and CMS performed searches in events containing one photon and
large missing transverse energy.  In this final state the dominant background process is 
$Z(\to \nu \nu)+\gamma$ production with smaller backgrounds coming from $W/Z+\gamma$ 
and $W/Z+\textrm{jet}$ production in which electrons or jets are mis-identified as photons. 

   The ATLAS collaboration has performed a mono-photon analysis using $4.7$~fb$^{-1}$ of $pp$ collisions at
 a center of mass energy of $\sqrt{s}=7$~TeV~\cite{Aad:2012fw}.
Events are required to pass an \etmiss~trigger with a threshold of 
$\etmiss>70$~GeV. Events are selected if they contain a photon candidate with $p_T(\gamma)~>~150$~GeV within a pseudo-rapidity of $|\eta|<2.37$, 
excluding the transition region between barrel and endcaps of the ATLAS detector 
($1.35<|\eta|<1.52$). These events are then required to contain $\etmiss>150$~GeV and not more than than one jet 
with $p_T(\textrm{jet})>30$~GeV  and $|\eta|<4.5$ are. Finally photons, jets and \etmiss~have to be well separated
 by $\Delta \phi>0.4$ between \etmiss and either photon or jet, and $\Delta R>0.4$ between photon and jet.

\begin{figure}[h!]
  \begin{minipage}{0.9\textwidth}
      \centering  
      \includegraphics[scale=0.6]{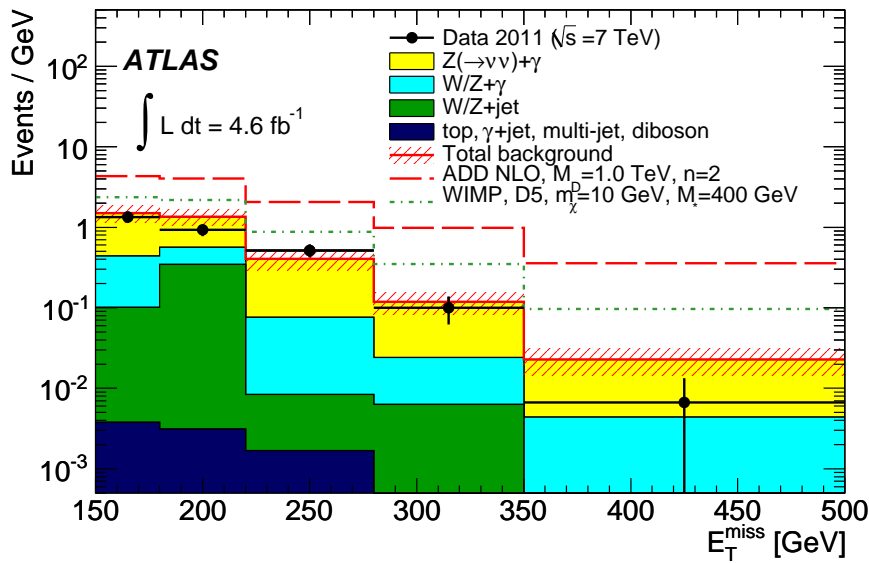}
   \end{minipage}
   \caption{\etmiss~ distribution observed after full event selection in the 
ATLAS mono-photon search~\cite{Aad:2012fw}.} 
    \label{fig:atlas_mp}
\end{figure}

Backgrounds are modeled using data and cross-checked in control regions. 
Multijet background production is estimated using data control regions.  The smaller 
background processes of single-top, diboson, $\gamma\gamma$, 
$\gamma+\textrm{jet}$ production are modeled using Monte Carlo simulations.  
Figure~\ref{fig:atlas_mp} shows the \etmiss~after full signal selection. 
No significant excess over the Standard Model expectation is observed.

   The CMS collaboration also performed a mono-photon analysis using $5.0$~fb$^{-1}$ 
of $pp$ collisions at a center of mass energy of $\sqrt{s}=7$~TeV~\cite{Chatrchyan:2012tea}.
The analysis uses single photon triggers which are fully efficient for the 
selected signal. The photons are required to fulfill $p_T(\gamma)>145$~GeV and to 
be within the central electromagnetic calorimeter $|\eta|<1.44$. The missing transverse
energy of these events has to exceed $\etmiss>130$~GeV. Events with a muon present or significant 
hadronic activity, signified by either having a track with 
$p_T>20$~GeV or a jet with $p_T(\textrm{jet})>40$~GeV within $|\eta|<3.0$ and 
$\Delta R>0.5$ of the photon axis, are rejected.

  The CMS collaboration has recently update this search ~\cite{CMS-PAS-EXO-12-047} with 19.6$fb^{-1}$ of data collected at $\sqrt{s}=8$~TeV. The analysis is similar except 
few differences: events are require to have  $p_T(\gamma)>145$~GeV and $\etmiss>140$~GeV, jet veto is optimized with one jet is allowed to be present in the event below $p_T(\textrm{jet})<30$~GeV. These jets are counted only if they qualify the pile-up jet identification criteria and have $p_T(\textrm{jet})>20$~GeV within $|\eta|<3.0$. The lepton veto is used where events are rejected if a lepton is present in the event with $p_{T}^{l}>10$~GeV and also satisfies the stringent lepton identification requirements. The SM prediction with all these selections is also tested using a control region where lepton veto are reversed to select a phase space dominated by SM electroweak backgrounds.

\begin{figure}[h!]
  \begin{minipage}{0.49\textwidth}
      \centering 
      \includegraphics[scale=0.42]{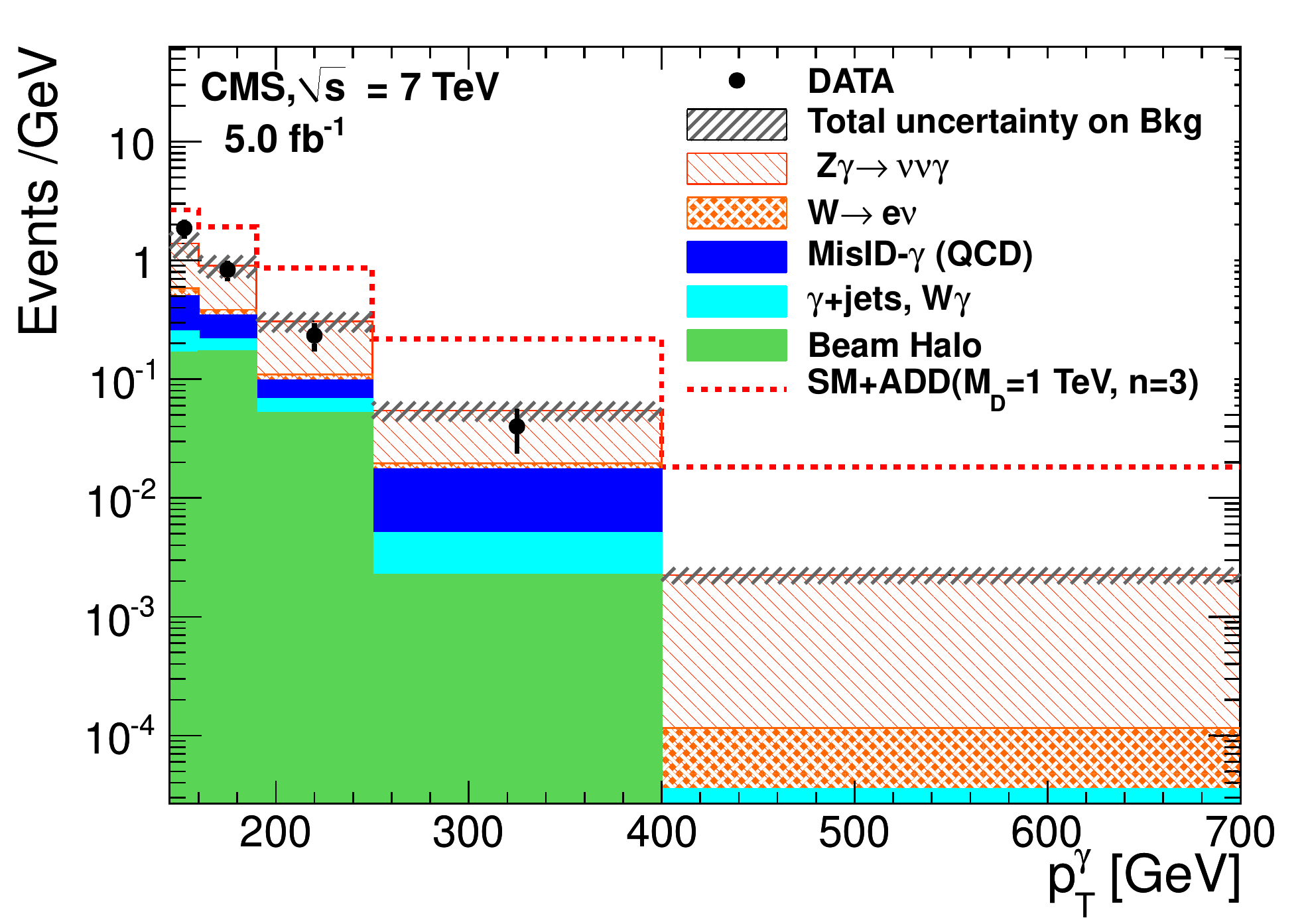}
      \end{minipage}
   \hfill   
  \begin{minipage}{0.49\textwidth}                           
      \centering          
      \includegraphics[scale=0.40]{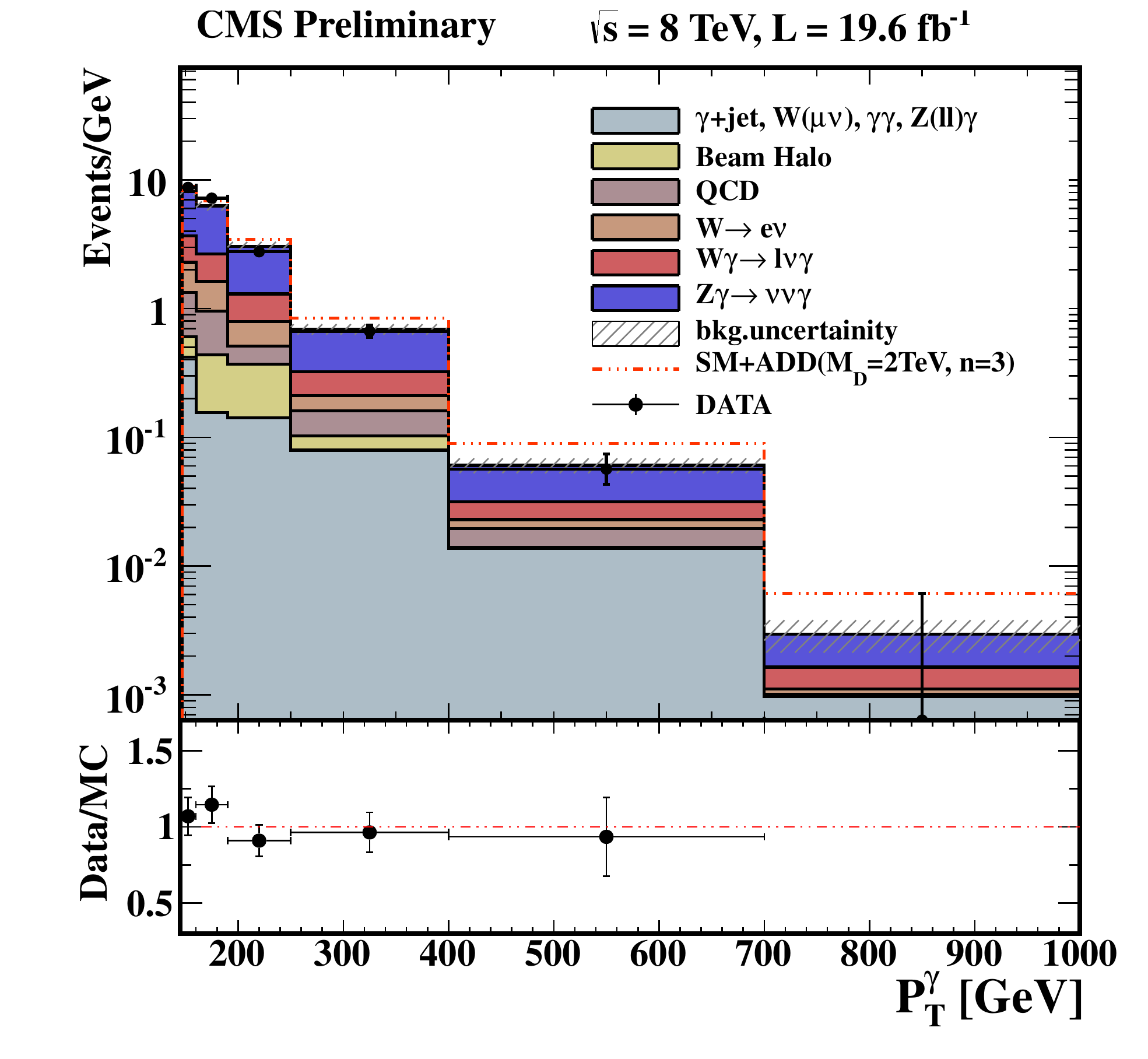}      
      \end{minipage}
   \caption{$p_T(\gamma)$ distribution after full selection in the CMS 
mono-photon analysis:(left) at $\sqrt{s}=7$~TeV using 5.0$fb^{-1}$ of data ~\cite{Chatrchyan:2012tea}, (right) at $\sqrt{s}=8$~TeV using 19.6$fb^{-1}$ of data ~\cite{CMS-PAS-EXO-12-047}. It should be noted that no WIMP signal contribution is 
plotted but the red dashed signal corresponds to
a particular model for large extra dimensions.~\cite{Chatrchyan:2012tea}}
    \label{fig:cms_mp} 
\end{figure}

 In both version of this search, instrumental backgrounds from electrons or jets reconstructed as photons are estimated using
data in control regions.  Additional ``out of time backgrounds'' from cosmic muons and beam halo 
are estimated using timing requirements, and the remaining backgrounds are estimated using 
simulation. Figure~\ref{fig:cms_mp} shows the corresponding $p_T(\gamma)$ spectrum from $\sqrt{s}=7$~TeV and $\sqrt{s}=8$~TeV searches. The $\sqrt{s}=7$~ TeV analysis estimate a cut-and-count based limits while in 8 TeV analysis shapes are exploited to get the cross section limits.  

\begin{figure}[h!]
  \begin{minipage}{0.9\textwidth}
    \centering
    \includegraphics[scale=0.65]{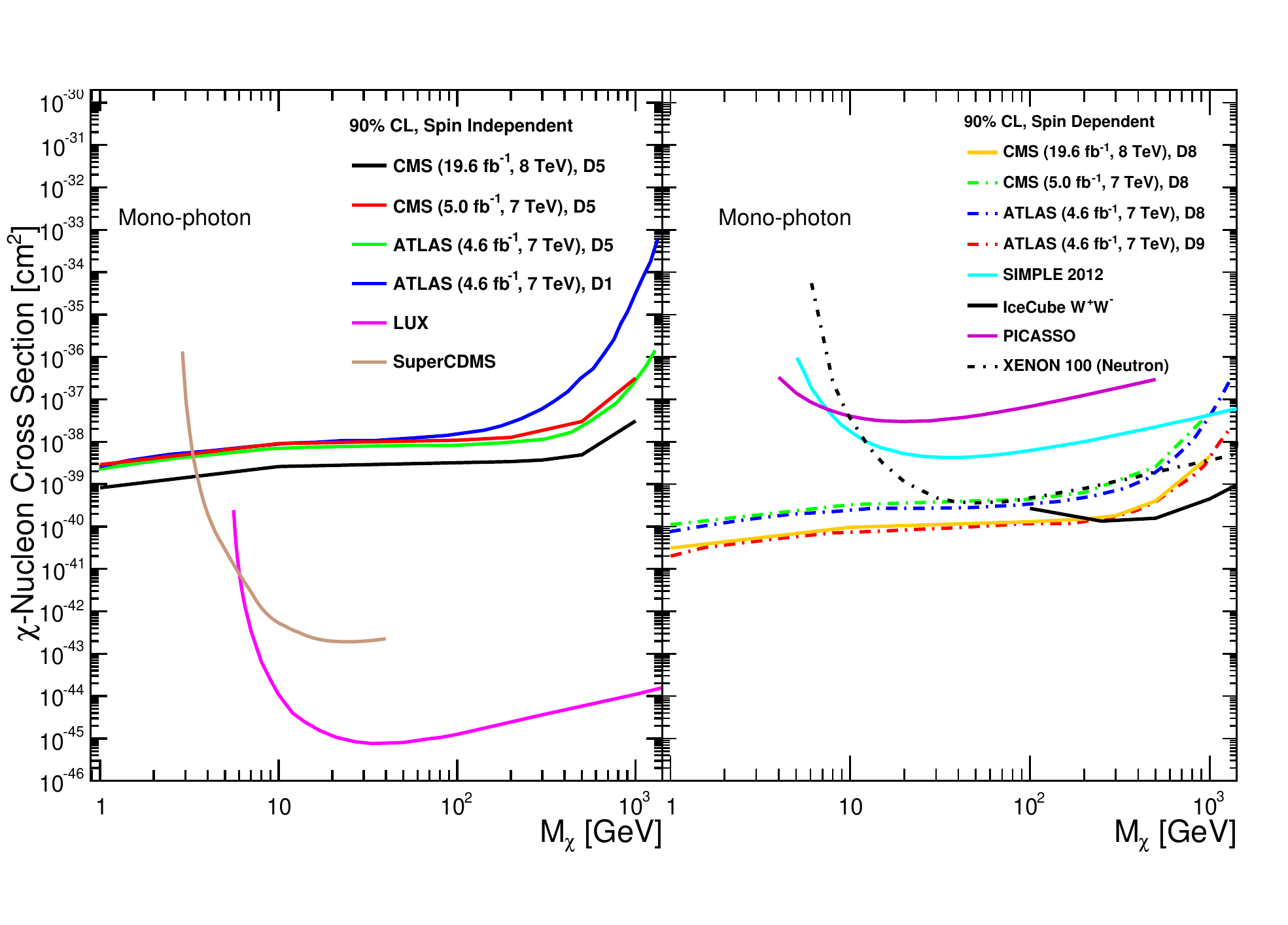}
  \end{minipage}
  \caption{Observed 90$\%$ CL upper limit for spin-independent(left) and 
spin-dependent(right) DM-nucleon cross section for mono-photon final state from 
the ATLAS and the CMS collaboration.  The results are also compared with latest 
direct detection bound from LUX ~\cite{Akerib:2013tjd}, SuperCDMS 
~\cite{Agnese:2014aze}, XENON-100 ~\cite{Aprile:2013doa}, IceCube 
~\cite{Aartsen:2012kia}, PICASSO~\cite{Archambault:2012pm}, and SIMPLE 
~\cite{Felizardo:2011uw}.}
  \label{fig:monoPho_search}
\end{figure}

  The ATLAS and the CMS collaborations have derived 90$\%$ CL upper limit 
on WIMP-nucleon cross section in the mono-photon final search. 
Figure~\ref{fig:monoPho_search} summarizes these results and comparisons 
with direct detection limits. For the spin-independent case, CMS considers 
the D5 operator while ATLAS uses D5 and the D1 operators. In the 
spin-dependent case both collaborations use D8 while ATLAS additionally studies the D9 operator. For spin-independent scenarios both 
collaborations obtain similar upper limits at $\sqrt{s}=7$~TeV while CMS results at $\sqrt{s}=8$~TeV provides the leading bound on WIMP-nucleon cross section across the  
entire DM mass range of $M_{\chi}= $1 - 1000 GeV. In comparison to direct searches from 
LUX~\cite{Akerib:2013tjd} and SuperCDMS~\cite{Agnese:2014aze} these limits improve only existing constraints for $M_{\chi}<$3 GeV, 
while for $M_{\chi}>$ 3 GeV direct detection  limits provide stronger bounds. In case of a spin-dependent coupling, the 
D9 results from ATLAS and D8 results from CMS provide similar and leading bounds up to $M_{\chi}$ of 200~GeV of, while for even larger $M_{\chi}$ IceCube reaches best sensitivity.
Bounds on the D8 operators also achieve stronger constraints to up to to about $M_\chi\sim100$ GeV compared to direct detection
experiments. It should be noted that direct detection limits from other experiments like, PICASSO, 
SIMPLE, XENON-100 provide most sensitive bounds on WIMP-nucleon cross section for various mass intervals in $M_{\chi}$ starting from a few GeV to 100 GeV, but generally these bounds are weakend in the spin-dependent case.

\subsection{Mono-jet searches}
ATLAS and CMS performed searches in the final state with a jet and
missing transverse energy. The main backgrounds to this analysis are $Z\to \nu \nu$, $W+\textrm{jets}$, 
single top, top pair, diboson and multijet events, whereas $Z\to \nu \nu$ is an irreducible background.

The ATLAS mono-jet analysis was performed using a luminosity of $4.7$~fb$^{-1}$ 
of data recorded at $\sqrt{s}=7$~TeV in 2011 and updated with $10.5$~fb$^{-1}$ 
at a center of mass energy of $\sqrt{s}=8$~TeV recorded in 
2012~\cite{ATLAS:2012ky, ATLAS-CONF-2012-147}. Both analyses utilize similar event 
selections. Events are required to pass an ~\etmiss based trigger of at least $\etmiss >$80~GeV, 
with an efficiency of about $95$\% for $\etmiss>120$~GeV. All events are required to pass the latter \etmiss~ criterion.
Events are also required to contain  one jet with $p_T(\textrm{jet})>120$~GeV and $|\eta|<2.0$. Events with more than 
two additional jet with $p_T(\textrm{jet}) > 30$~GeV and 
$|\eta|<4.5$ are rejected. Multijet events which may pass the event selection due to the 
mis-measurement of one of the jets gives rise to (fake)~\etmiss. Those are suppressed 
by ensuring that the sub-leading jet is well separated from the direction of 
$\etmiss$.  $W/Z$ production is  suppressed by applying a veto on electrons and muons.

\begin{figure}[h!]
  \begin{minipage}{0.49\textwidth}
      \centering 
      \includegraphics[scale=0.42]{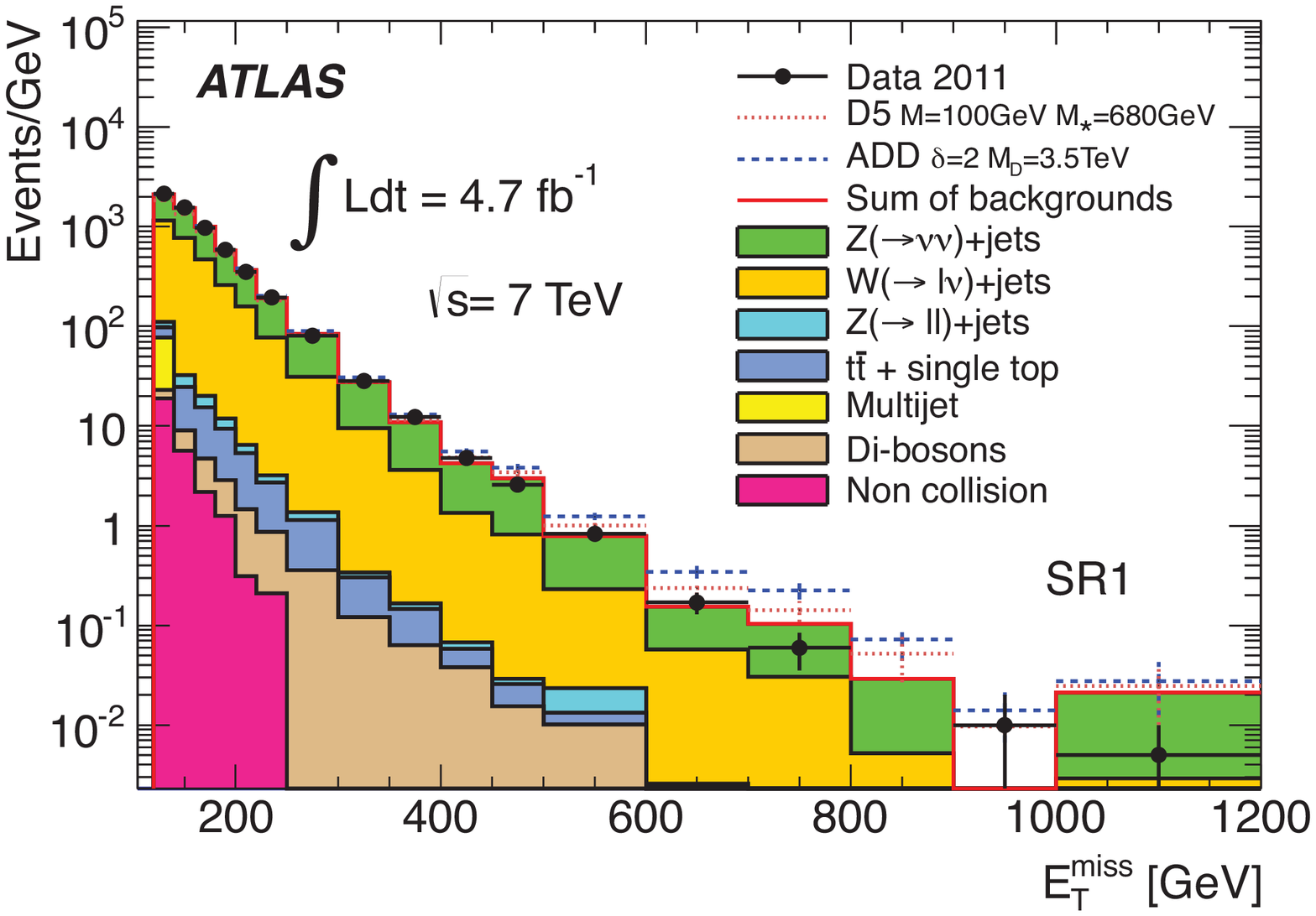}
      \end{minipage}
    \hfill
    \begin{minipage}{0.49\textwidth}
      \centering 
      \includegraphics[scale=0.42]{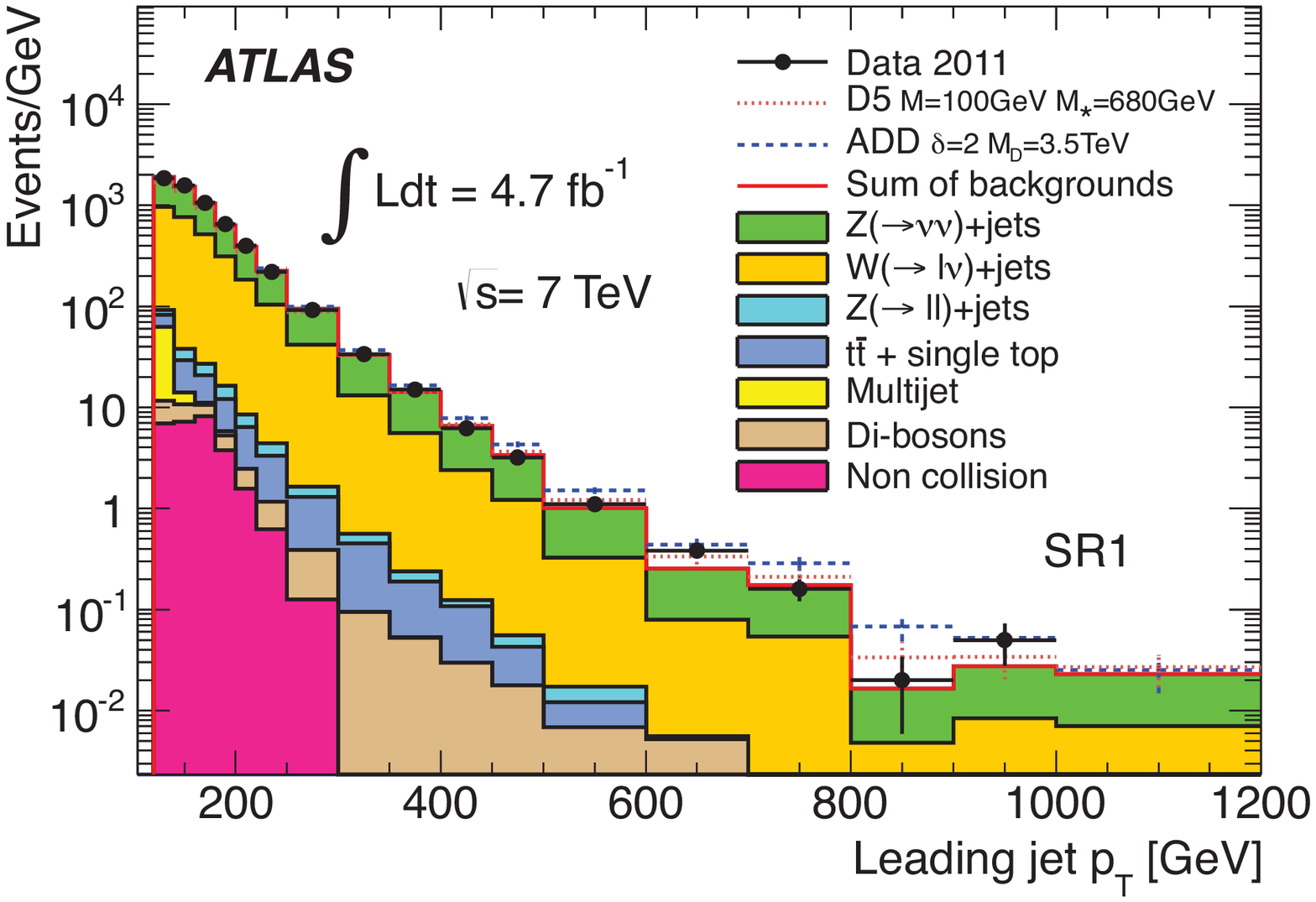}
    \end{minipage}
   \caption{\etmiss~(left) and transverse momentum of the leading jet (right) in 
loosest signal region of the ATLAS monojet search~\cite{ATLAS:2012ky, 
ATLAS-CONF-2012-147}. 
    \label{fig:atlas_mj} }
\end{figure}

Background from  $Z\to \nu \nu$ is estimated using a $Z\to \ell \ell$ control region and a 
transfer functions to account for kinematic differences of background contributions in the hadronic final 
state. A similar procedure is employed for $W+\textrm{jets}$ production. Top and 
diboson backgrounds are taken from Monte Carlo simulations. Multijet production is estimated 
from data by an enriched selection of this process. Figure~\ref{fig:atlas_mj} 
shows \etmiss~ and leading jet $p_T$ of the signal region requiring 
$\etmiss,~p_T(\textrm{jet})>120$~GeV. The search is performed in 4 signal 
regions, defined by lower bounds on the leading jet momentum and \etmiss with 
values of $120,~220,~350~\textrm{and}~500$~GeV, respectively. None of the signal
regions show any significant excess over expected background.

The CMS mono-jet analysis was performed using $4.7$~fb$^{-1}$ of $\sqrt{s}=7$~TeV data and 
$19.5$~fb$^{-1}$ of $\sqrt{s}=8$~TeV ~\cite{Chatrchyan:2012me, 
CMS-PAS-EXO-12-048}. The data used to study events with a single jet and missing transverse energy 
are collected using a combination of jet and \etmiss~ triggers. The \etmiss~ 
trigger uses a threshold of $80~(120)$~GeV for the $7~(8)$~TeV analysis.  All 
triggers are fully efficient in the selected signal regions. Events are selected by 
requiring $\etmiss>200$~GeV and that the momentum of 
the leading jet exceeds $p_T(\textrm{jet})>110$~GeV. Events containing a second jet with 
$p_T(\textrm{jet})>30$~GeV are accepted, but any event with more jets of 
$p_T(\textrm{jet})>30$~GeV are rejected. To suppress multijet background 
the angular separation between leading and sub-leading jets has to be 
less than $\Delta \phi(j_1, j_2)<2.5$ and also leptons with reconstructed leptons. In contrast to the ATLAS
analysis even with isolated tracks $p_T(trk)>10$~GeV are also rejected. All backgrounds except $Z\to 
\nu\nu+\textrm{jets}$ and $W+\textrm{jets}$ are modeled using MC simulation. These
data driven backgrounds are modeled using signal regions dominated by $W \to \mu +\textrm{jet}$ and cross checked
using the statistically limited $Z\to \mu \mu$ process for $Z \to \nu \nu$.

\begin{figure}[h!]
    \begin{minipage}{0.49\textwidth}
      \centering 
      \includegraphics[scale=0.42]{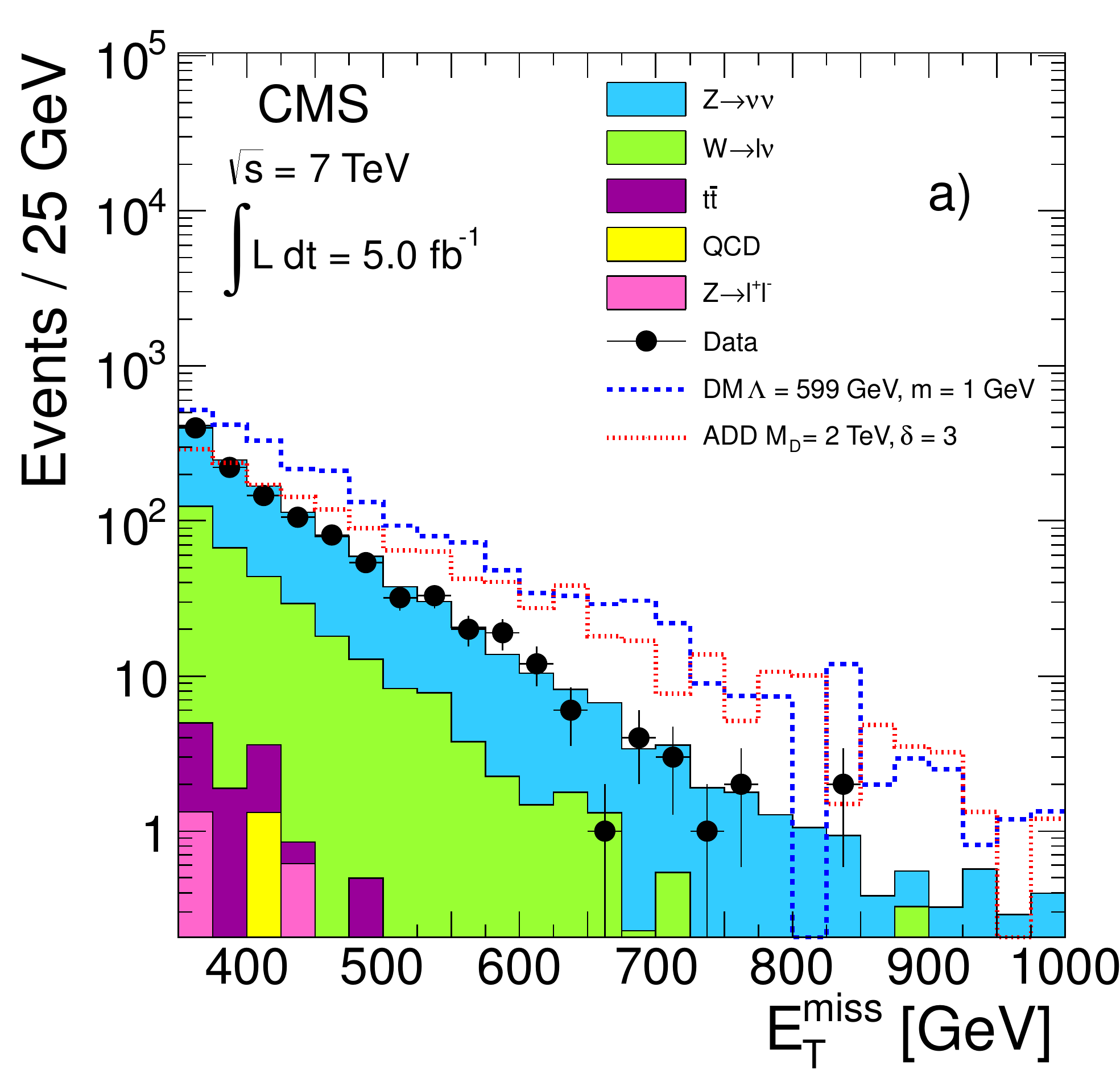}
    \end{minipage}
    \begin{minipage}{0.49\textwidth}
      \centering 
      \includegraphics[scale=0.40]{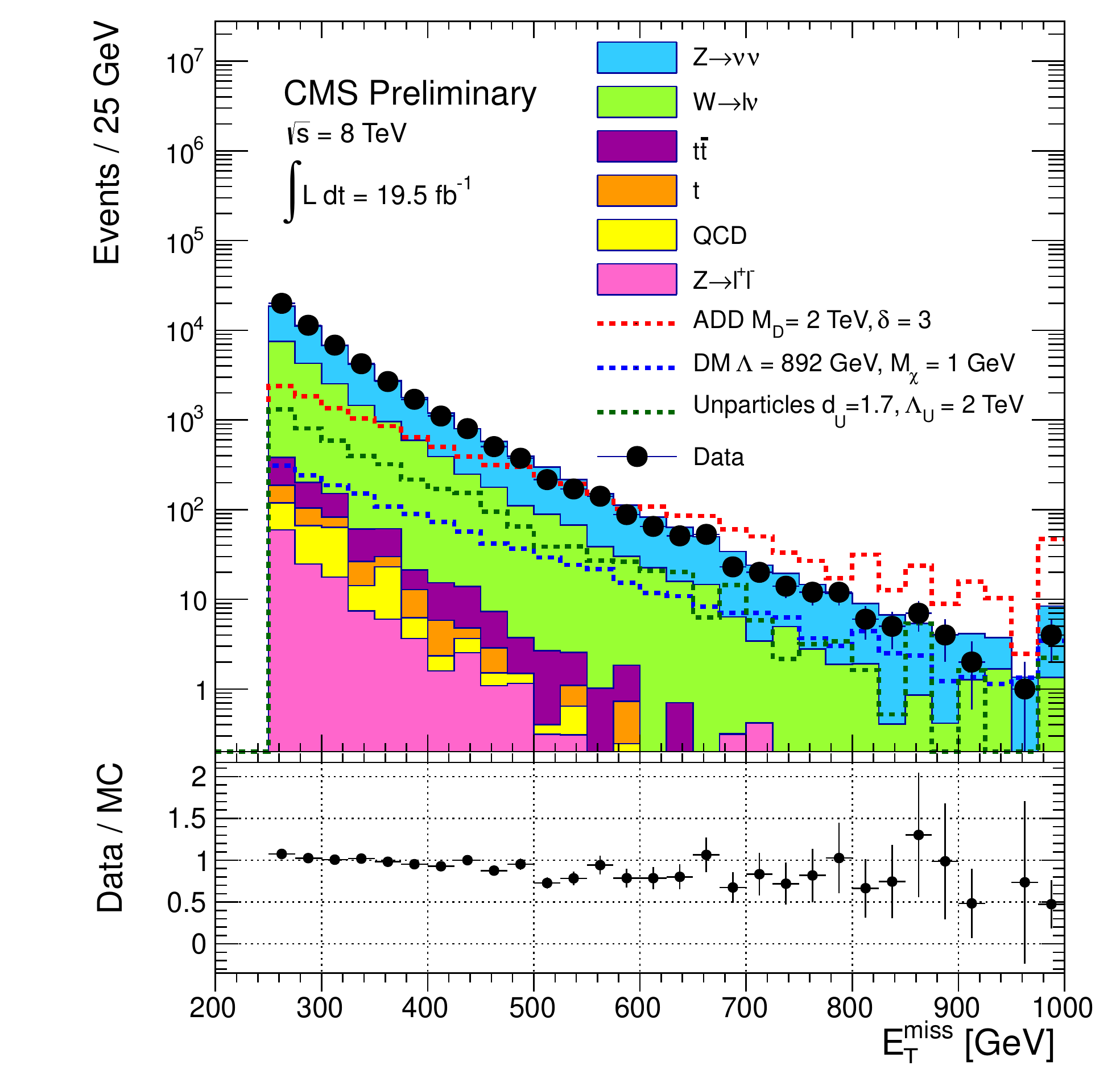}
    \end{minipage}
   \caption{\etmiss~distribution in the signal region of the CMS monojet 
analysis for $\sqrt{s}=7$ (left) and $\sqrt{s}=8$ (right). A simulated dark 
matter signal for axial-vector couplings and $m_{\chi}=1$~GeV is shown as dashed 
blue line.~\cite{Chatrchyan:2012me, CMS-PAS-EXO-12-048}}
   \label{fig:cms_mj}
\end{figure}

Various signal regions in ~\etmiss between $\etmiss>250-550$~GeV are 
studied. Final \etmiss selection for DM production is $\etmiss>350~(400)$~GeV in 
the 7~TeV (8~TeV) analysis. The \etmiss~distributions for both analyses 
are shown in Figure ~\ref{fig:cms_mj}.

   Mono-jet searches at LHC have been performed at $\sqrt{s}=$7 and 8 TeV and 
might be expected to provide the strongest bounds on WIMP-nucleon cross section 
due to the fact that the rate of gluon or quark initial-state radiation is large 
relative to photon, $W$ or $Z$ boson radiation. The systematic uncertainties 
associated with this final state, however, are more severe than for more 
clean channels. Both the CMS and the ATLAS collaboration have considered D5 and 
D11 operators for spin-independent, and the D8 operator for spin-dependent 
coupling. In addition ATLAS has evaluated limits for D9 which are slightly 
stronger bounds on the scattering cross section. For the spin-independent 
couplings CMS results for D11 present the best upper limits for $M_{\chi}<$ 10 
GeV while for rest of $M_{\chi}$ masses LUX has stronger upper bounds. CMS also 
has the leading collider bounds on mono-jets from the D5 operator, stronger than 
direct detection searches for $M_\chi<3$ GeV.
\begin{figure}[h!]

  \begin{minipage}{0.9\textwidth}
    \centering
    \includegraphics[scale=0.65]{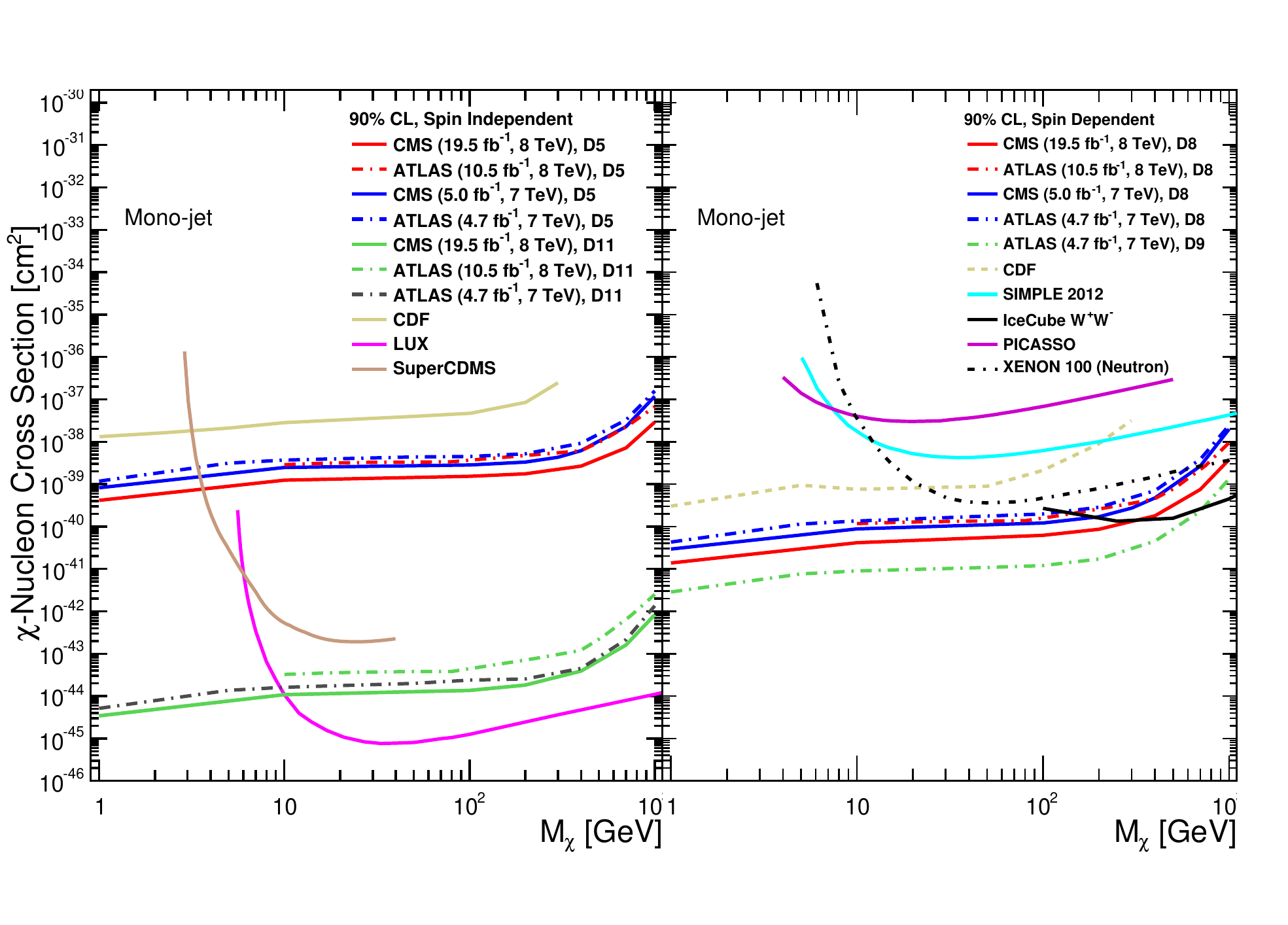}
  \end{minipage}
  \caption{Observed 90$\%$ CL upper limit for spin-independent(left) and 
spin-dependent(right) DM-nucleon cross section for mono-jet final state from the 
ATLAS, CMS, and CDF~\cite{Aaltonen:2012jb} collaboration. The results are also 
compared with latest direct detection bounds from LUX~\cite{Akerib:2013tjd}, 
SuperCDMS~\cite{Agnese:2014aze}, XENON-100~\cite{Aprile:2013doa}, 
IceCube~\cite{Aartsen:2012kia}, PICASSO~\cite{Archambault:2012pm}, and 
SIMPLE~\cite{Felizardo:2011uw}.}
  \label{fig:monoJet_search}
\end{figure}
 
\begin{figure}[h!]                                                               
  \begin{minipage}{0.65\textwidth}                                               
    \centering                                                                   
    \includegraphics[scale=0.55]{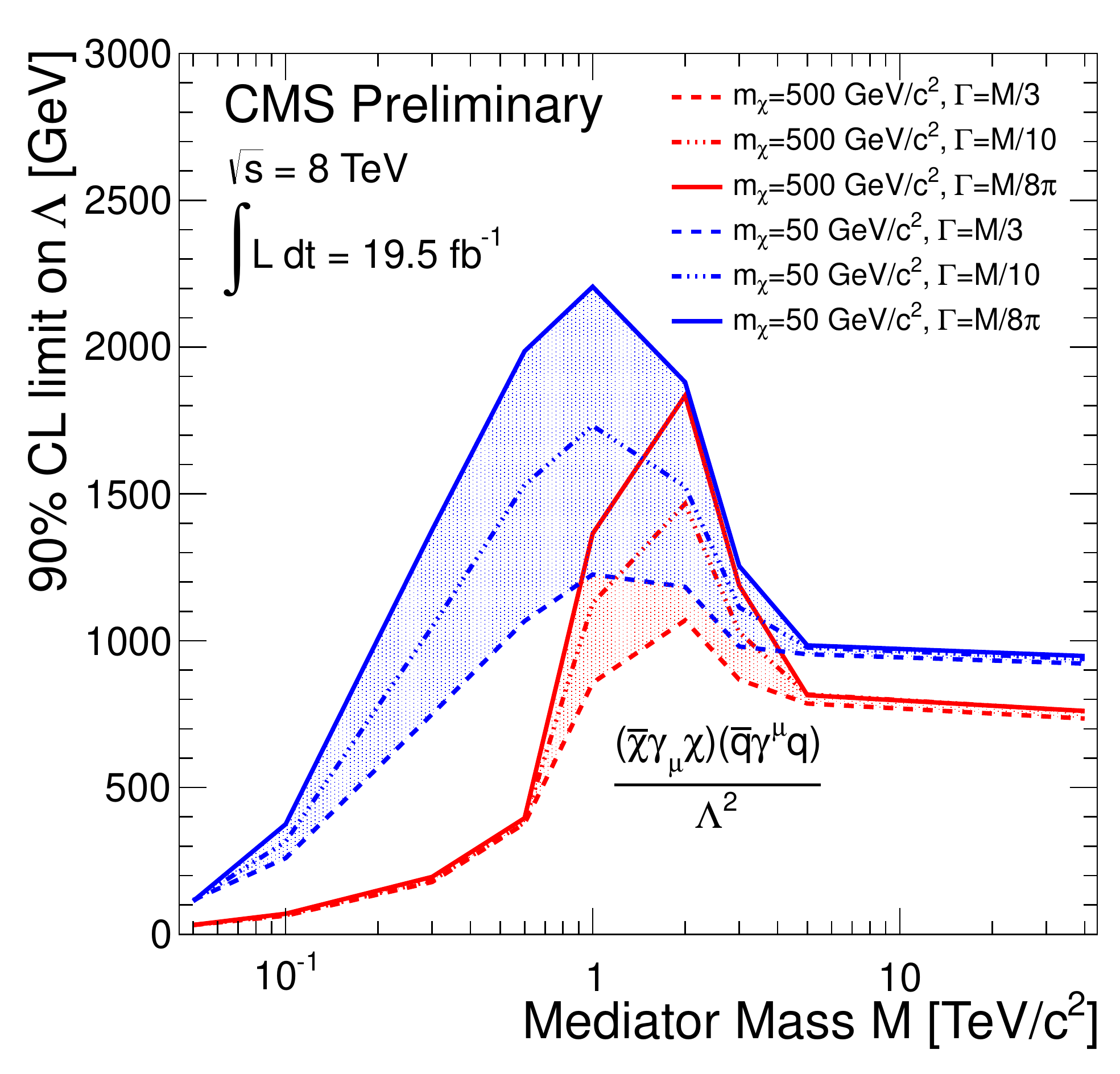}
  \end{minipage}                                                                 
  \caption{90$\%$ CL observed limit on $\Lambda$ as a function of mass of the 
mediator(M) for two different $M_{\chi}$ values of 50 GeV and 500 GeV. The 
widths are also varied from M/3 and M/8$\pi$.}
  \label{fig:medScan} 
\end{figure}

   As discussed in Sec.~\ref{sec:effth} it is generally assumed  that limits 
obtained with collider-accessible mediator mass are weaker than the effective 
field theory. CMS has also shown results with limits obtained for different 
mediator mass (M) and widths ($\Gamma$). These results are shown in 
Figure~\ref{fig:medScan}. As the mass of the mediator approaches the kinematic 
range, the production cross section shows resonant enhancement and indicates 
on-shell production. As expected, when the mediator is light compared to the 
kinematic scale of the events, the resulting bounds grow weak quickly.  The 
limits obtained for the vector couplings from these variations in mass and 
widths show that for heavy mediator mass they are approximately same as obtained 
from effective theory (EFT) framework ~\cite{Fox:2011pm,Buchmueller:2013dya}. For mediator mass 
range from few hundreds of GeV to few TeV, EFT limits are weaker due to an 
enhancement in the production cross section. In the region below a few hundreds
GeV these limits are too strong. Figure~\ref{fig:medScan} also shows that limits 
for heavy mediator are stronger relatively at lower values of $M_{\chi}$ which 
is complementary to the direct detection limits.
 
\begin{figure}[h!]
\vspace{-26mm}
  \begin{minipage}{0.65\textwidth}
   \centering
    \includegraphics[scale=0.55]{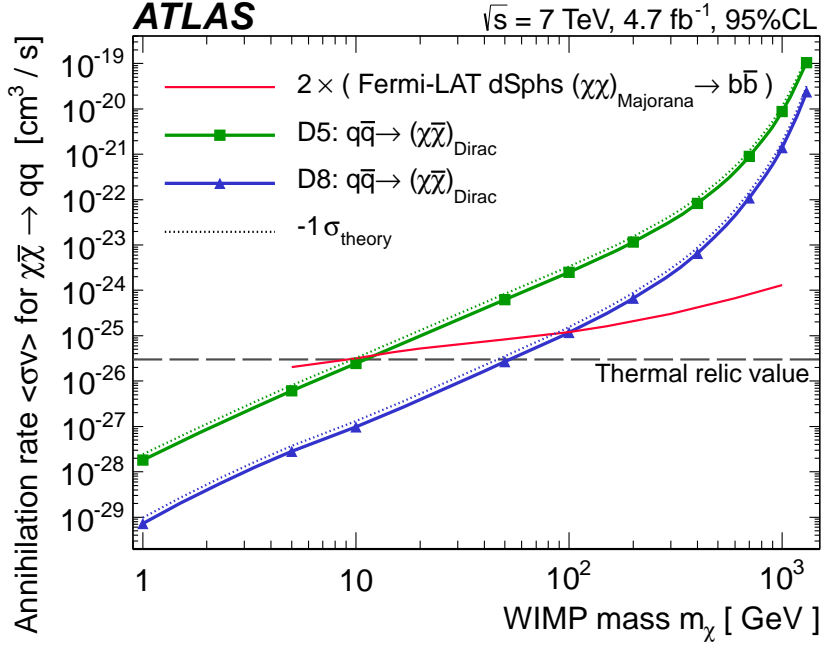}
  \end{minipage}
  \caption{90$\%$ CL observed limit on WIMP annihilation rates $<\sigma v >$ as 
a function of $M_{\chi}$(solid line). The impact of theoretical uncertainties on 
observed limits is shown as dotted lines. }  
\label{fig:annihilationRate}                                                     
\end{figure}

   The lower limits on $\Lambda$ from collider searches can also be expressed in 
terms of relic abundance of WIMP production as described in 
Ref.~\cite{Fox:2011pm, Beltran:2010ww}. The ATLAS collaboration has translated 
their results of the mono-jet search at $\sqrt{s}=$ 7 TeV to WIMP annihilation 
rate into four light flavors of quark assuming equal coupling strength for all 
of them. These annihilation rates are estimated by translating vector and 
axial-vector limits as described in~\cite{Fox:2011pm}. The annihilation rate is 
defined as $<\sigma v>$ where $\sigma$ is the cross section and $v$ is the 
average relative velocity of dark matter. The limits are based on the assumption 
of 100$\%$ branching ratio of WIMPs to four light flavour of quarks. These 
results are summarized in Fig.~\ref{fig:annihilationRate}. The figure also shows 
a comparison with the observations of Galactic satellite galaxies with the 
Fermi-LAT experiment~\cite{Ackermann:2011wa} for Majorana WIMPs. For WIMPs to 
make up the relic abundance, annihilation rate must be above the thermal relic 
value observed by WMAP( dashed line)~\cite{Komatsu:2010fb}.

\subsection{Mono-$W$/$Z$ searches}
The ATLAS collaboration has performed searches for DM production with $W$ or 
$Z$ boson in their final state, probing either the hadronic or the leptonic 
decay mode of the $Z$.  The CMS experiment also performed a search for DM production
with the $W$ boson decaying leptonically.

\subsubsection{$W/Z \to \textrm{jet}$ + \etmiss}

The ATLAS collaboration is searching for dark matter pair production 
in association with a $W$ or $Z$ decaying hadronically using $20.3$~fb$^{-1}$ 
of data at $\sqrt{s}=8$~TeV ~\cite{Aad:2013oja}. In contrast to other searches presented 
that assume equal couplings of the dark matter particles to up-type and down-type quarks, as they 
are not sensitive to a relative sign between those two couplings. For $W$ boson 
radiation there is interference between the diagrams in which the $W$ boson is 
radiated from the $u$ quark or the $d$ quark. In the case of equal couplings, the 
interference is destructive and gives a small $W$ boson emission rate. However, 
if the up-type and down-type couplings are of opposite sign, one finds constructive 
interference and $W$ boson emission becomes the dominant process, even over 
radiation of photons or gluons.

This search uses particularly wide jets using the {\it 
Cambridge-Aachen-Algorithm}~\cite{Dokshitzer:1997in} with wide jet radii of 
$R=1.2$. These jets are intended to capture the full decay products of 
hadronically decaying $W$ and $Z$ bosons. The internal structure of these 
wide-radius jets is probed in terms of the momentum balance of their leading two 
constituents and the jet mass is calculated. Are are reconstructed using the $anti-k_T$ algorithm
using $R=0.4$.

Events are recorded using a $\etmiss$ trigger that is fully efficient at the 
minimum requirement of $\etmiss>150$~GeV. Events have to contain 
at least one large-radius jet of $p_T(\textrm{jet})>250$~GeV and 
$|\eta(\textrm{jet})|<1.2$. The jet mass should be compatible with a $W$ or $Z$ 
boson, $50<m(\textrm{jet})<120$~GeV. $\sqrt{y}>0.4$ is required to suppress 
backgrounds from non-hadronic $W$ or $Z$ boson decays. Two signal regions are 
defined using $\etmiss>350~\textrm{and}~500$~GeV. Events with more than one 
narrow jet of $p_T>40$~GeV are rejected if the narrow jets do not overlap with 
the wide radius jet or if any narrow jet has $\Delta \phi (\etmiss, j)<0.4$. Any 
event containing a reconstructed photon or lepton (electron or muon) candidate 
with $p_T>10$~GeV in the fiducial detector region are rejected as well.

Dominant backgrounds to this selection are $Z\to\nu\nu + \textrm{jet}$ 
production and $W/Z+\textrm{jets}$ production with leptonic decays in which the 
lepton was not reconstructed. These backgrounds are estimated using data control 
regions with an inverted muon veto. Diboson, single- and double-top production 
are taken from simulations. Multijet production is negligible as background. 

\begin{figure}[h!]
    \begin{minipage}{0.99\textwidth}
      \centering 
      \includegraphics[scale=0.6]{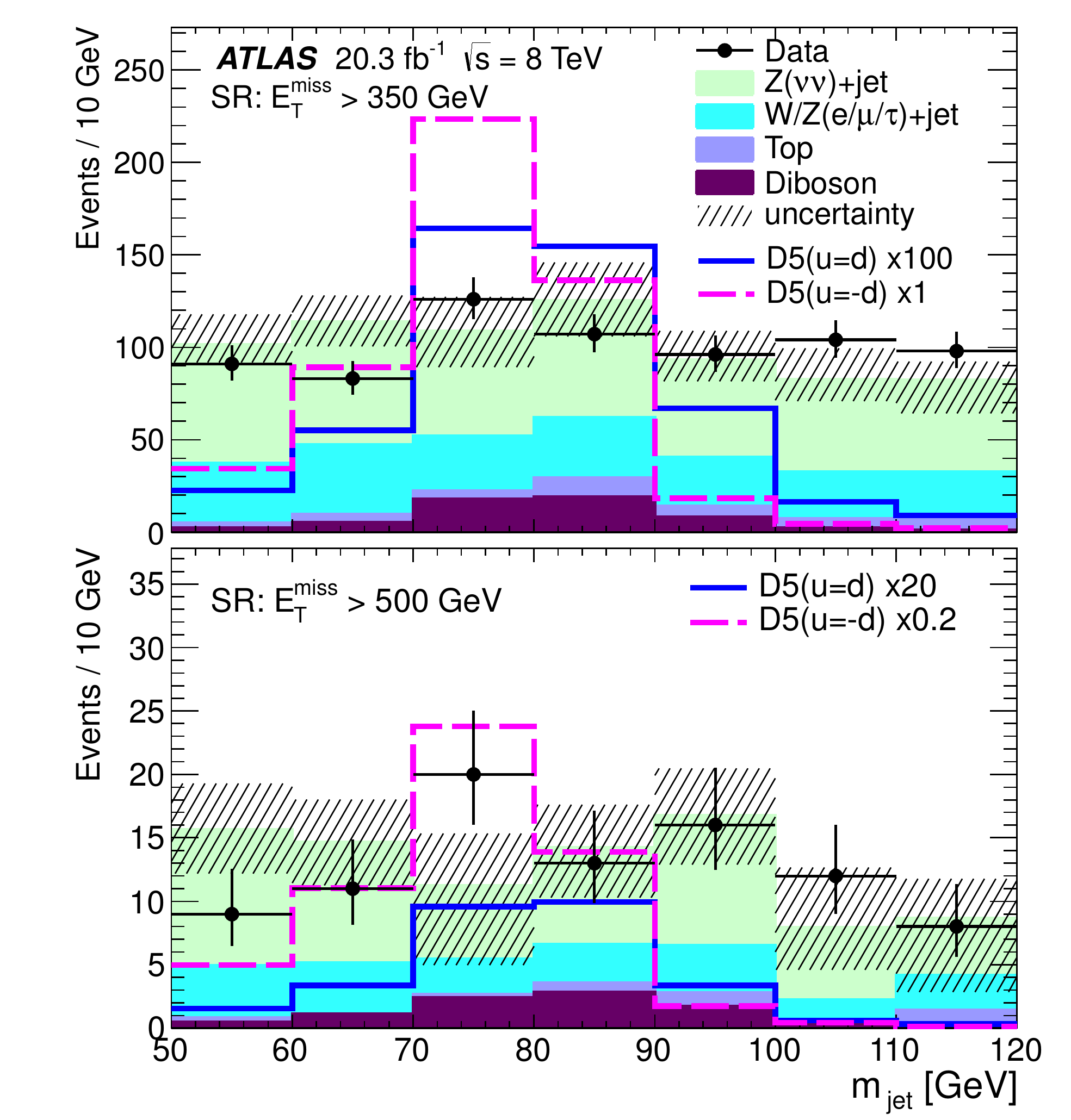}
    \end{minipage}
    \caption{Distribution of $m(\textrm{jet})$ in data and for the predicted 
background in the signal regions (SR) with $\etmiss>350$~GeV (top) and 
$\etmiss>500$~GeV (bottom). Also shown are the combined mono-$W$-boson and 
mono-$Z$-boson signal distributions with $m_{\chi}=1$~GeV and $M_*=1$~TeV for 
the $D5$ destructive and $D5$ constructive cases, scaled by factors defined in 
the legends. Uncertainties include statistical and systematic contributions. 
\cite{Aad:2013oja}}
   \label{fig:atlas_dijet}
\end{figure}

After full selection no significant excess is observed for either signal region.
Figure~\ref{fig:atlas_dijet} shows the $m(\textrm{jet})$ distribution 
for both signal regions.

\subsubsection{$W\to \ell \nu$ + \etmiss}

A search for dark matter production in final states with an electron or a muon 
and a neutrino has been performed by the CMS collaboration using $20$~fb$^{-1}$ 
of data recorded at $\sqrt{s}=8$~TeV~\cite{CMS-PAS-EXO-13-004}. The results can 
be interpreted in terms of the cross section of a $W$-boson recoiling against a 
pair of dark matter particles in an effective theory, considering vector- and 
axial-vector like couplings. Also varying coupling strength to up and down 
type quarks is assumed, parametrized by $\xi$ with $\xi = 0,\pm 1$. 

Candidate events are recorded using single-lepton trigger. The reconstructed 
momentum required for muons is $p_T(\mu)>45$~GeV and for electrons is 
$p_T(e)>100$~GeV. Further criteria are applied by requiring 
$0.4<p_T(\ell)/\etmiss<1.5$ and an angular separation between lepton and 
\etmiss~ of $\Delta \phi(\ell,\nu)>0.8\pi$. The main observable in this analysis 
is the transverse mass of the lepton-\etmiss system:

\begin{equation}
  M_T=\sqrt{2\cdot p_T(\ell)\cdot \etmiss \cdot (1-\cos \Delta \phi(\ell,\nu)}
\end{equation}

\begin{figure}[h!]
  \begin{minipage}{0.49\textwidth}
    \centering
    \includegraphics[scale=0.42]{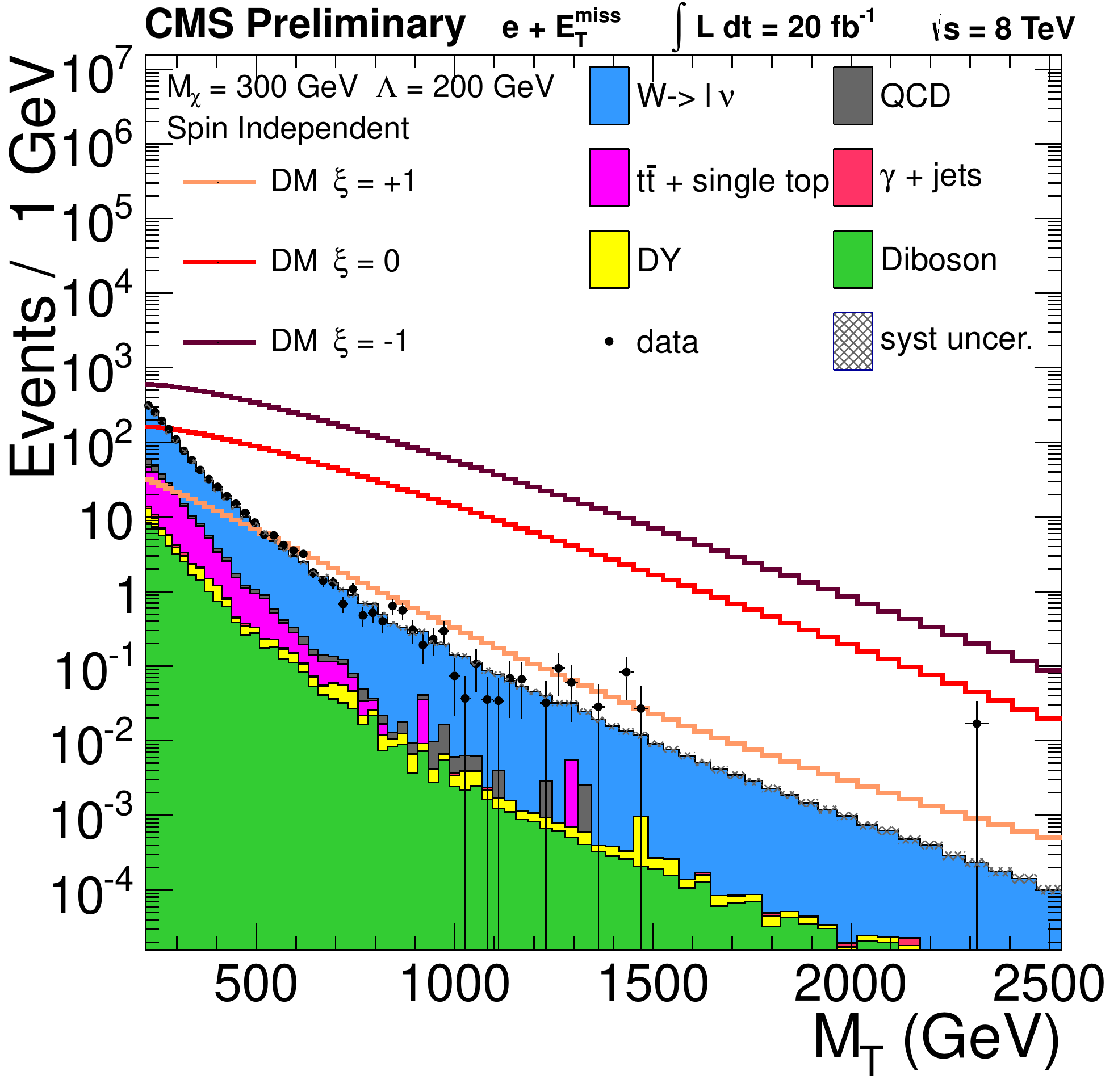}
  \end{minipage}
  \begin{minipage}{0.49\textwidth}
    \centering
    \includegraphics[scale=0.42]{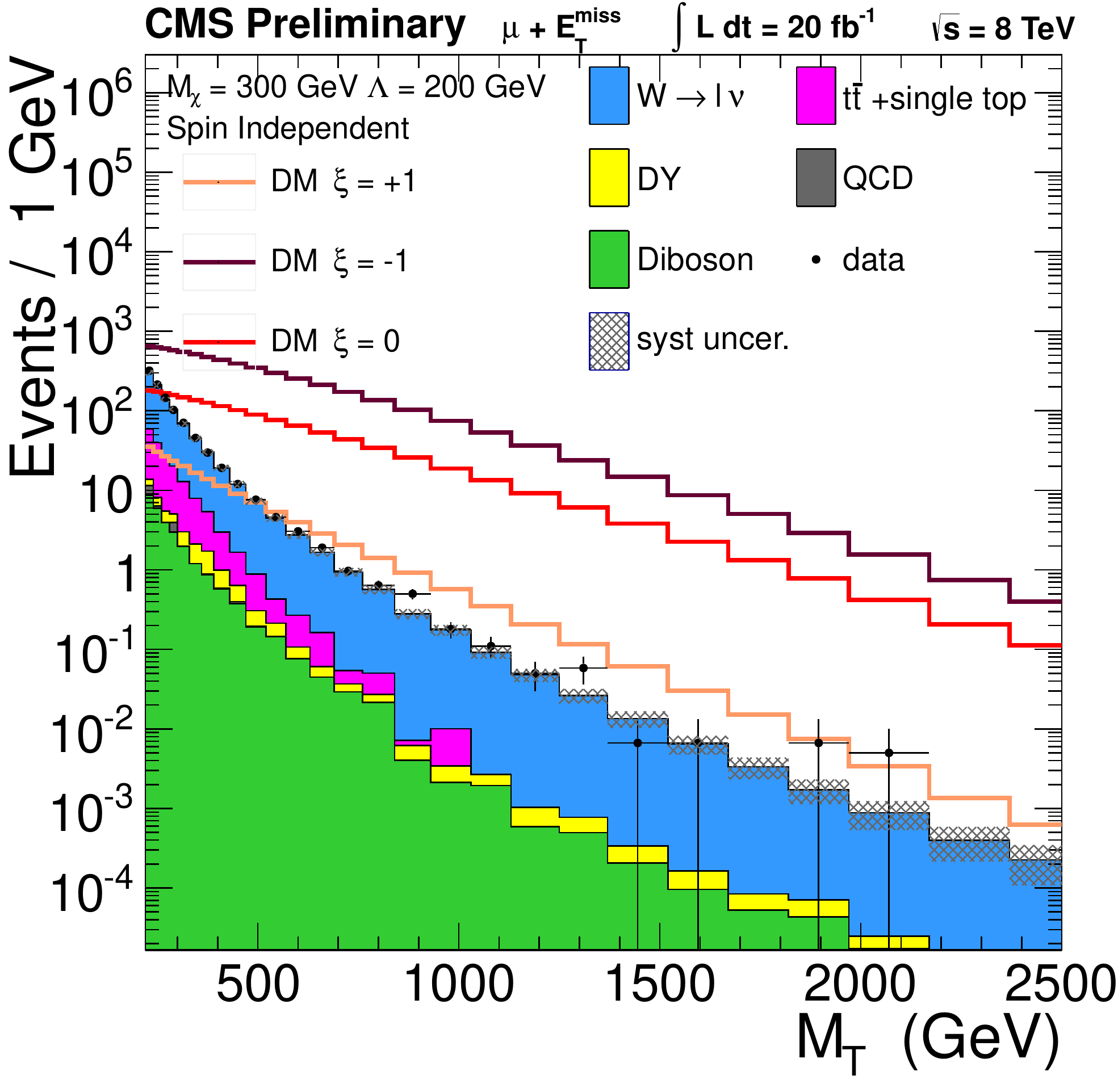}
  \end{minipage}
  \caption{  Transverse mass distribution for the electron (left) and muon 
(right) channel. The shaded area represents the total systematic uncertainty on 
the SM background. Simulated signal distributions for a Mχ =300 GeV, Λ=200 GeV 
and ξ = ±1, 0 are also shown. \cite{CMS-PAS-EXO-13-004}}
  \label{fig:cms_monolepton}
\end{figure}

The dominant background is the high transverse mass tail of the $W\to \ell \nu$ 
decay. Further background arise from multijet, top-quark, $Z$ boson and diboson 
production. All backgrounds are modeled using simulations. Due to low statistics 
at very high $M_T$  the full $M_T$ distribution is fit using an empirical 
function and extrapolated to the
region of interest. After full selection again no significant deviation between 
background expectation and data is observed. Figure~\ref{fig:cms_monolepton} 
shows the $M_T$ distributions for the electron and muon channels along with 
different couplings for up/down type quarks for DM production.

\subsubsection{$Z\to \ell \ell$ + \etmiss}

The search for dark matter production in association with a $Z$ boson decaying 
to two leptons has been performed by the ATLAS collaboration using $20.3$fb$^{-1}$ of data collected at 
$\sqrt{s}=8$~TeV~\cite{Aad:2014vka}. The data were recorded using a combination 
of single and di-lepton triggers. Events were selected if they contained two 
electrons or muons with $p_T(\ell)>20$~GeV  and invariant mass 
$76<m(\ell\ell)<106$~GeV forming a $Z$ boson candidate. To suppress events with 
\etmiss from mis-measured jets $\Delta \phi (\etmiss, p_T(\ell \ell))>2.5$ is 
required. The absolute value of the pseudorapidity of the dilepton system is 
required to be $\eta(\ell \ell)<2.5$, and
$|p_T(\ell \ell)~-~\etmiss|/p_T(\ell \ell)<0.5$, where $p_T(\ell \ell)$ is the 
transverse momentum of the dilepton system. Events are removed if they contain a 
jet with $p_T(\textrm(jet))>25$~GeV or a third lepton with $p_T(\ell)>7$~GeV. 
Four inclusive signal regions with $\etmiss>150,~250,~350, \textrm{and}~400$~GeV 
are defined.

\begin{figure}[h!]
    \begin{minipage}{0.99\textwidth}
      \centering 
      \includegraphics[scale=0.7]{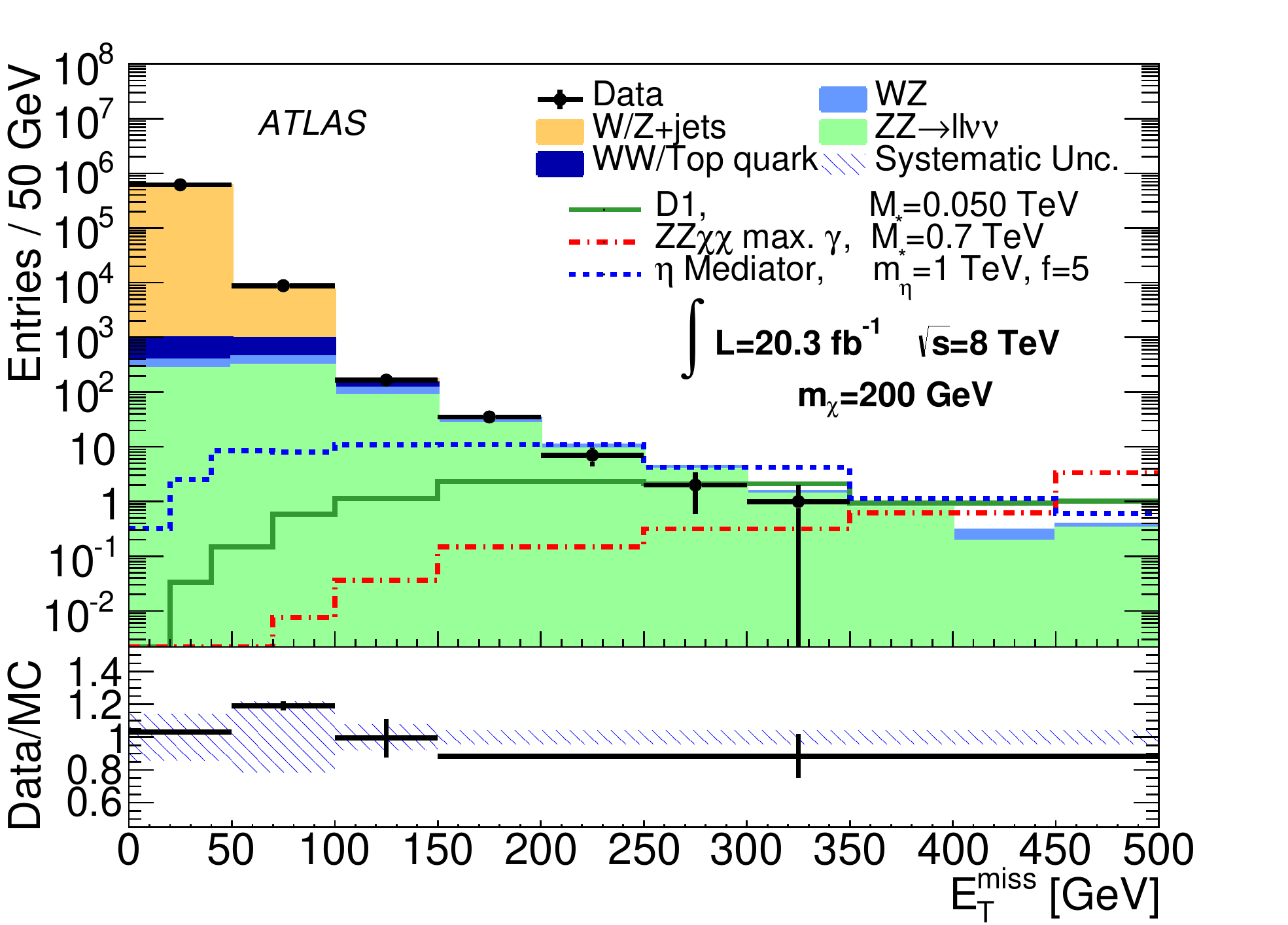}
    \end{minipage}
    \caption{\etmiss~distribution  of the ATLAS mono-$Z$ with $Z\to \ell \ell$ 
search~\cite{Aad:2014vka}.}
   \label{fig:atlas_dilepton}
\end{figure}

The dominant background is diboson production. $WZ$ and $ZZ$ are estimated using 
simulation; $WW$, $t\bar{t}$, $Wt$, $Z\to \tau \tau$ and $W/Z+\textrm{jets}$ are 
estimated from data. No significant excess over background 
expectation  is observed after full selection as seen in 
Fig.~\ref{fig:atlas_dilepton}.

\begin{figure}[h!]                                                               
  \begin{minipage}{0.9\textwidth}
    \centering
    \includegraphics[scale=0.65]{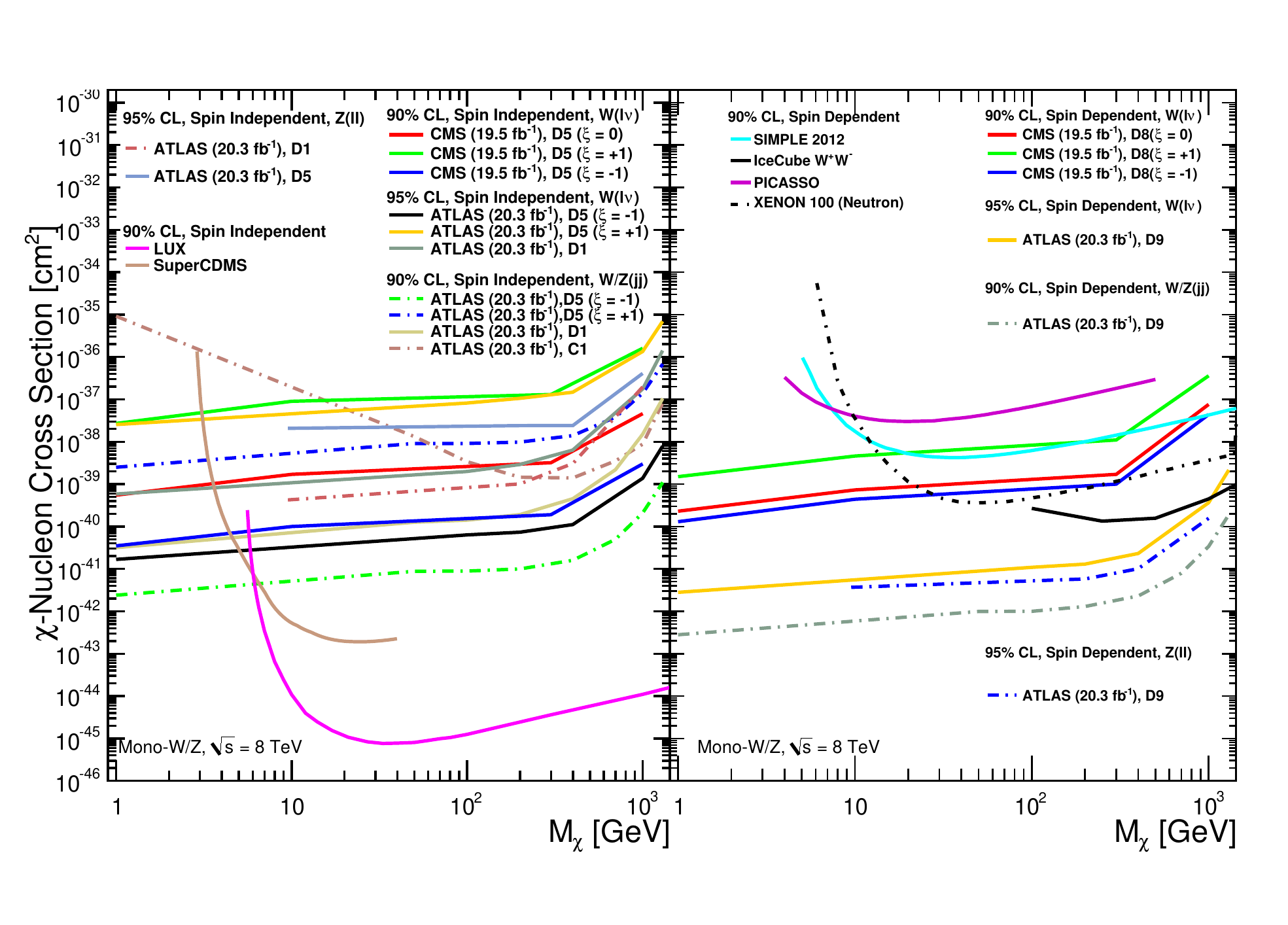}
  \end{minipage}
  \caption{Observed  upper limits for spin-independent(left) and 
    spin-dependent(right) DM-nucleon cross section for mono-W and mono-Z final state 
    from the ATLAS and the CMS experiments. The results are also compared with 
    latest direct detection bound from LUX~\cite{Akerib:2013tjd}, 
    SuperCDMS~\cite{Agnese:2014aze}, XENON-100~\cite{Aprile:2013doa}, 
    IceCube~\cite{Aartsen:2012kia}, PICASSO~\cite{Archambault:2012pm}, and 
    SIMPLE~\cite{Felizardo:2011uw}.}
  \label{fig:monoWZ_search}
\end{figure}

  The production of a DM pair in association with a $W$ or $Z$ boson is different 
with respect to the photon or jet cases as the sign of the DM coupling to 
up-type and down-type quarks play an important role in production rate, hence 
limits are also sensitive to these couplings. Figure~\ref{fig:monoWZ_search} 
summarizes these searches from the CMS and the ATLAS collaborations for 
different mono-W/Z final states.
  
  For $W\rightarrow l\nu$  decay the strongest observed bound for 
spin-independent and spin-dependent couplings are from the ATLAS collaboration 
with D5($\xi = -1$) and D9 operators.  It should be noted that the ATLAS limits are evaluated for 95$\%$ 
CL. The D1 case is much more weakly bounded, as are cases where the couplings to 
up- and down-type quarks do not lead to such strong constructive interference, 
while bounds on D8 comparable to those from spin-dependent direct detection.
  
 Similarly for $W/Z \rightarrow$jj decay mode D5($\xi = -1$) and D9 gives the 
best limits. The $Z \rightarrow\ell\ell$ final state is analyzed by the ATLAS 
collaboration with D1 and D9 giving the strongest observed 95$\%$ CL upper 
limits for spin-independent and spin-dependent couplings respectively. The most 
stringent bound among all the final states for mono-W/Z production are from 
hadronic final states. Direct search limits from LUX are the strongest current bounds
for $M_{\chi}>$ 6 GeV for any spin-independent scenario. The limits on the D9 
operators are stronger than the direct detection bounds for the entire 
$M_{\chi}$ mass range accessible to the LHC for spin-dependent interactions. 
  
\subsection{Constraints from Higgs boson measurements}

Various models predict that a `hidden' dark matter sector is coupled to the 
visible standard model via the Higgs sector, which allows for a renormalizable 
direct coupling. The Higgs boson then would decay into additional weakly 
interacting massive particles.  This would subsequently be observable as a 
deviation from the expected Standard Model Higgs branching ratios. Measurements 
of Higgs boson properties and dedicated searches performed by the ATLAS and CMS 
collaborations lead to constraints on the $H\to invisible$ branching ratio and 
therefore couplings to dark matter particles.  
\subsubsection{Limits for $ZH$ associated production}

Both the ATLAS and CMS collaborations have performed searches for the invisible 
decay of Higgs bosons using the associated production of a Higgs boson with a 
$Z$ boson.

\begin{figure}[h!]
  \begin{minipage}{0.8\textwidth}
    \centering
    \includegraphics[scale=0.6]{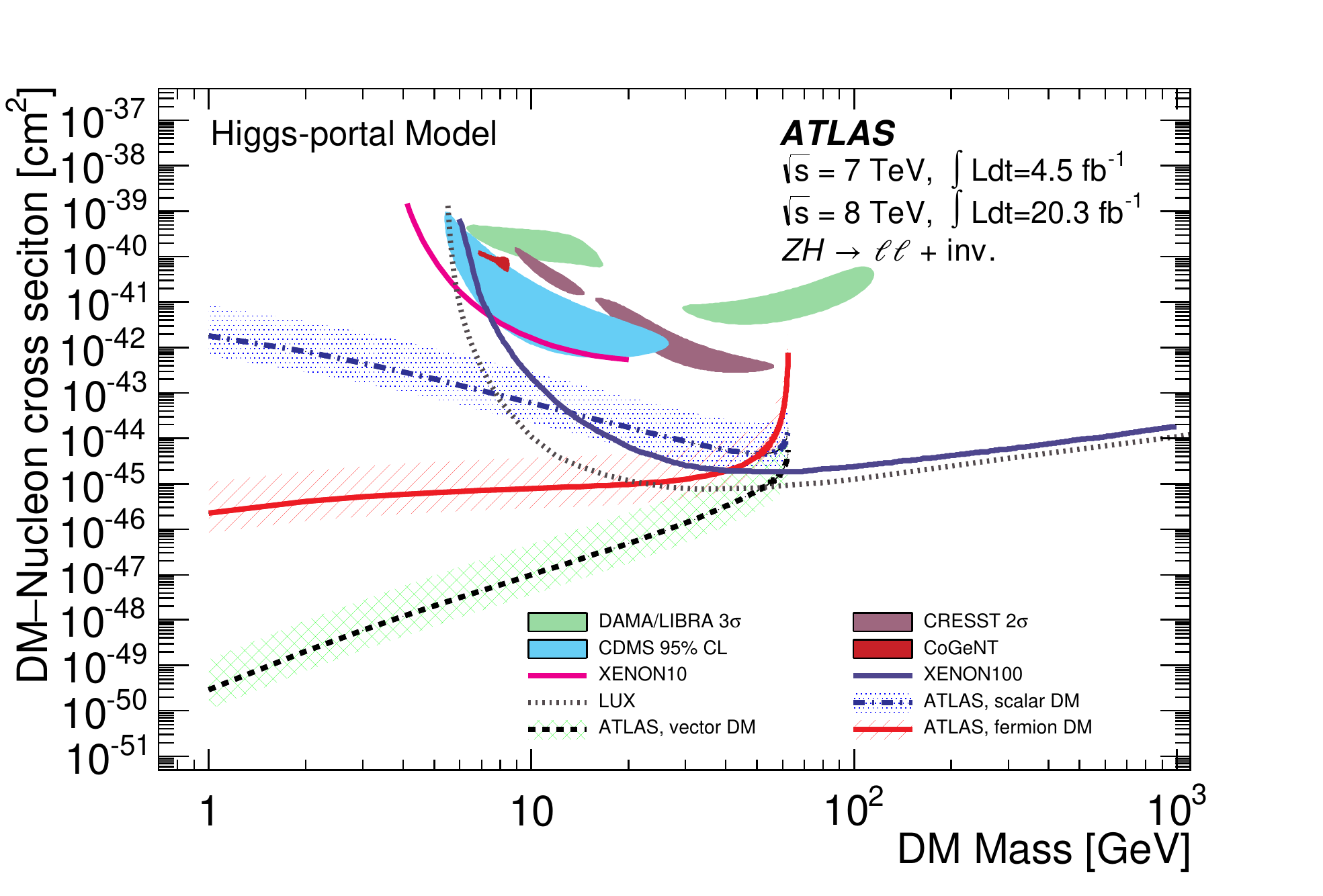}
  \end{minipage}
  \caption{ Limits on WIMP-nucleon scattering cross section at 90\% CL, 
extracted from the $BR(H \to~inv.)$ limit in a Higgs-portal scenario, compared 
to direct DM searches \cite{Aad:2014iia}}
  \label{fig:atlas_zh}
\end{figure}

ATLAS uses the entire 2011 $\sqrt{s}=7$~TeV ($4.7$~fb$^{-1}$) and 2012 
$\sqrt{s}=8$~TeV ($20.3$~fb$^{-1}$) datasets to search for anomalous decay of 
the Higgs boson in associated $ZH$ production~\cite{Aad:2014iia}. Events with a 
dilepton pair (electron or muons) with an invariant mass of $76 < m_{\ell 
\ell}<106$~GeV and $\etmiss>90$~GeV are selected. An additional robust version 
of the \etmiss is defined for this analysis:(\ptmiss~) which is the missing 
transverse momentum based on the reconstructed tracks.  In order to suppress 
events in which \etmiss~arises from mis-measurements in the calorimeter 
\etmiss~and \ptmiss~are required to point in the same direction with $\Delta 
\phi(\etmiss,\ptmiss)<0.2$.  The $Z$ and the invisible decaying Higgs boson are 
expected to form a back-to-back topology, therefore the azimuthal opening angle 
between \etmiss~and the dilepton system is required to be $\Delta \phi(p_T^{\ell 
\ell}, \etmiss)>2.6$ and the angle between the lepton pair is required to be 
less than $\Delta \phi(\ell,\ell)<1.7$ due to the boost of the dilepton system.
 The momentum of the reconstructed $Z$ boson should be balanced by the invisibly 
decaying Higgs boson so $|\etmiss-p_T^{\ell \ell}|/p_T^{\ell \ell}<0.2$ is 
required. A maximum likelihood fit to the \etmiss~distribution is performed and 
no significant excess is observed. Limits on the cross section times branching 
ratio for a Higgs boson decaying to invisible particles are set for a mass range 
of $110 < m_H < 400$~GeV and converted to the WIMP-nucleon scattering cross 
section assuming Higgs portal dark matter and $m_\chi \leq m_H/2$. 
Figure~\ref{fig:atlas_zh} presents the limits on the spin-independent 
WIMP-nucleon scattering cross section assuming Higgs portal dark matter.

The CMS collaboration searches for invisible decays of the Higgs boson using 
both associated $ZH$ production and vector boson fusion (VBF) production. The 
$ZH$ searches utilize final states with a pair of charged leptons (electrons or 
muons) or a b-quark pair ($b\bar{b}$)~\cite{Chatrchyan:2014tja}. Analyses are 
based on $4.9$~fb$^{-1}$ of $\sqrt{s}=7$~TeV data and $19.7$~fb$^{-1}$ of 
$\sqrt{s}=8$~TeV data. 

\begin{figure}[h!]
  \begin{minipage}{0.49\textwidth}
    \centering
    \includegraphics[scale=0.4]{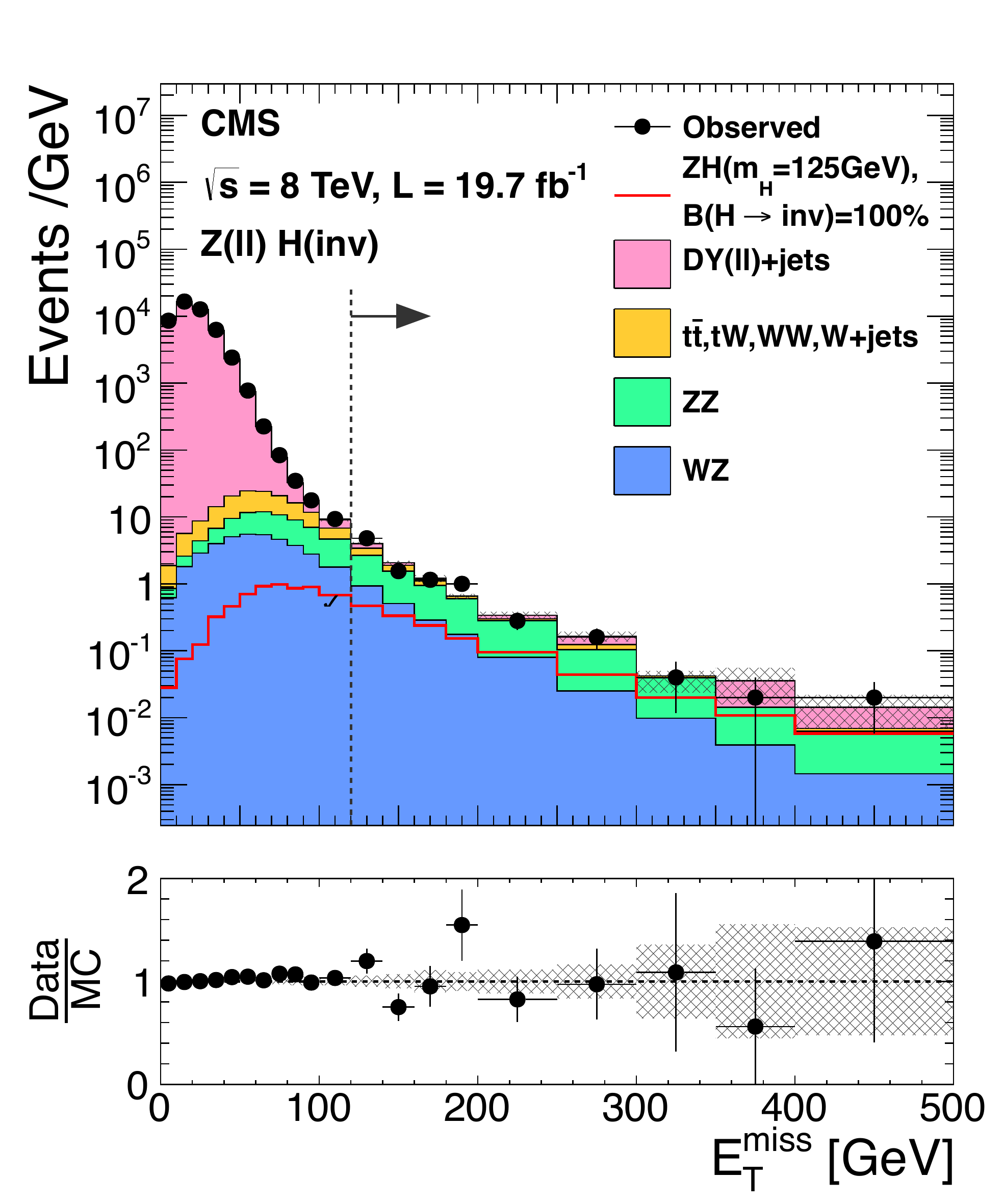}
  \end{minipage}
  \begin{minipage}{0.49\textwidth}
    \centering
    \includegraphics[scale=0.47]{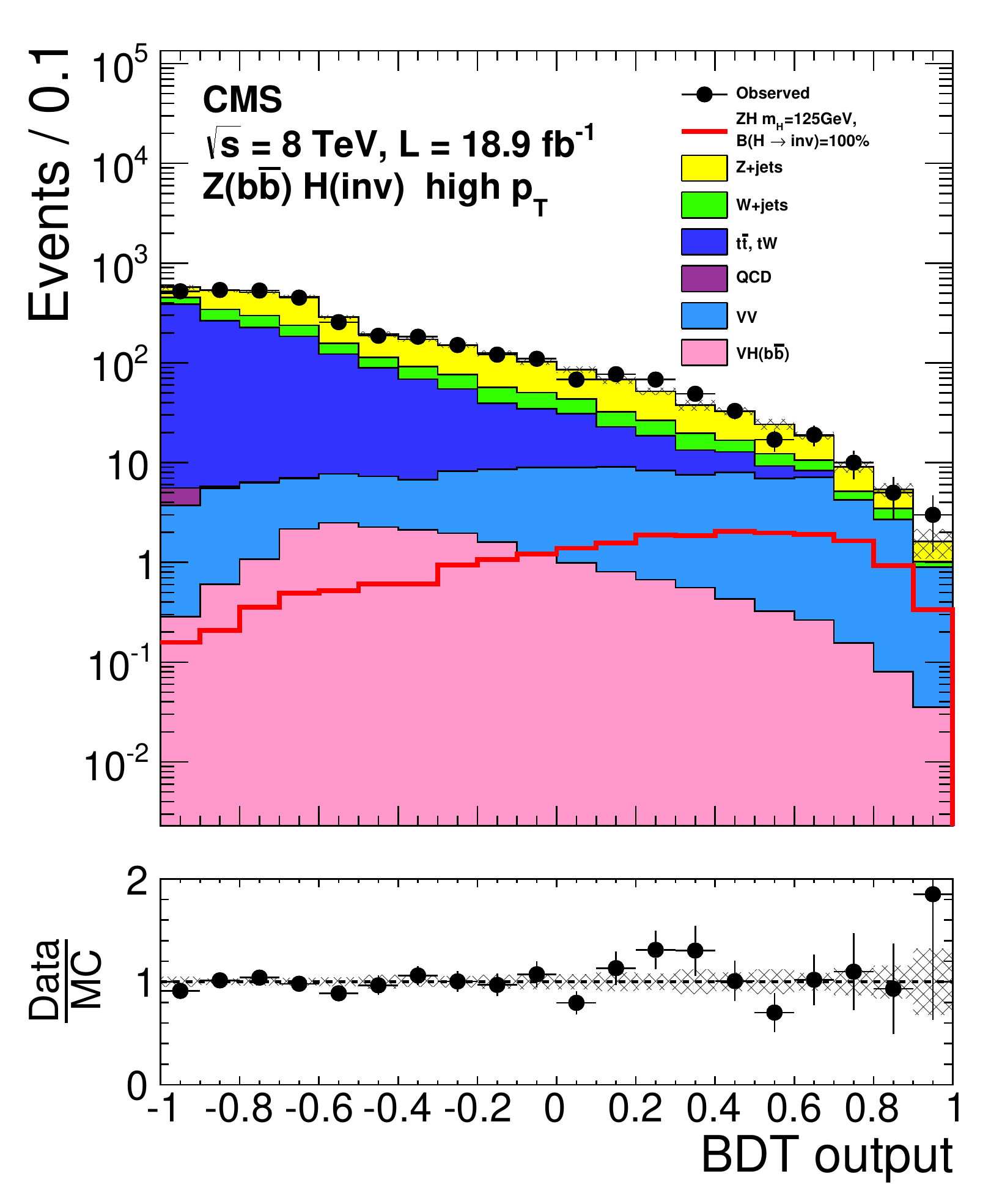}
  \end{minipage}
  \caption{Main discriminating variables in their final selection for the $Z(\to 
\ell \ell) H$ (left) and $Z(\to b\bar{b}) H$ (right, high $pT$ region is shown) 
analyses.~\cite{Chatrchyan:2014tja}}
  \label{fig:CMS_zh}
\end{figure}

\paragraph{Associated $ZH$ production}
 The dilepton channel requires a pair of isolated leptons with an invariant mass 
within $\pm 15\%$ of the mass of the $Z$ boson. Events with any $b$-tagged jet or more 
than one light flavor jet are rejected. In the ATLAS analysis similar requirements are 
applied with $\etmiss>120$~GeV,  $\Delta \phi(p_T^{\ell \ell}, 
\etmiss)>2.7$ and  $|\etmiss-p_T^{\ell \ell}|/p_T^{\ell \ell}<0.25$ to further 
increase the selection purity. The~\etmiss distribution obtained using 
the $Z(\to \ell \ell)H(\to inv.)$ selection is shown in Fig.~\ref{fig:CMS_zh}.

The $Z(\to b\bar{b})H(\to inv.)$ uses an event selection requiring large 
$\etmiss$ and a $b$-quark pair consistent with the $Z$ boson mass. Events are 
recorded using $\etmiss$ and $\etmiss+\textrm{jets}$ triggers. 
In these events \etmiss~and \ptmiss~ are required to point into the same 
direction $\Delta \phi(\etmiss,\ptmiss) < 0.5$. The analysis is further divided 
into three regions, based on \etmiss, denoted ``low'' ($100<~\etmiss<~130$~GeV), 
``intermediate'' ($130<\etmiss<170$~GeV), and ``high'' ($\etmiss>170$~GeV). The 
azimuthal separation between \etmiss~ and the closest jet is required to be 
$\Delta \phi(\etmiss,j)>0.5$ for the ``high'' region and for ``low'' and 
``intermediate'' regions $\Delta \phi(\etmiss,j)>0.7$ is required. 

For the low mass region, the \etmiss~significance, the ratio of $\etmiss$~ and 
the square root of the scalar sum of transverse energy of all particle-flow 
objects, is required to be greater than 3~GeV$^\frac{1}{2}$. Events with 
isolated leptons are rejected and the two jets forming the $Z$-candidate are 
required to be $b$-tagged.
The analysis then uses a boosted decision tree (BDT) discriminator to separate 
signal candidates from background events. The BDT is trained using simulated 
samples for signal and all background processes after the full selection 
described above. This is performed separately for each Higgs boson mass 
considered. 
Figure~\ref{fig:CMS_zh} shows the \etmiss~ and BDT output for $ZH$ analyses. 

\begin{figure}[h!]
  \begin{minipage}{0.49\textwidth}
    \centering
    \includegraphics[scale=0.55]{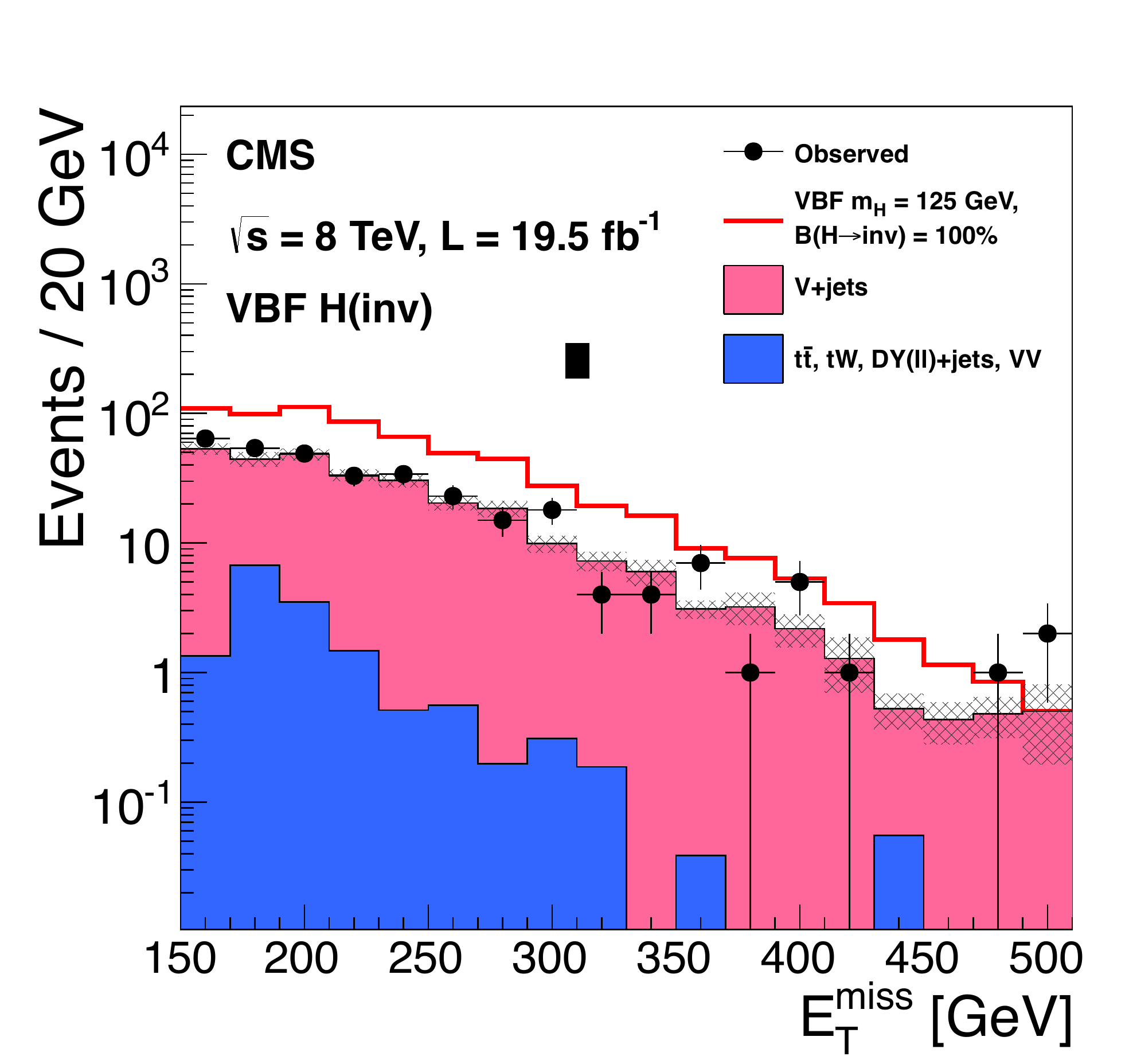}
  \end{minipage}
  \caption{\etmiss~distribution after full selection of the VBF 
analysis.~\cite{Aad:2014iia}}
  \label{fig:CMS_vbf}
\end{figure}

\paragraph{Vector Boson Fusion}
The VBF topology is distinguished by two jets in opposite forward directions. 
The events are recorded with a dedicated trigger for this topology, requiring 
$\etmiss>65$~GeV, requiring the transverse momentum of the forward jets to be 
$p_T>40$~GeV, and requiring that the invariant mass of the forward jets be 
$M(jj)>800$~GeV.  This selection is further tightened offline by requiring 
$p_T(\textrm{jets})>50$~GeV, $|\eta(\textrm{jet})|<4.7$, $\eta(j_1) \cdot 
\eta(j_2)<0$, $\Delta\eta(j_1,j_2)>4.2$, $\etmiss>130$~GeV, and 
$M(jj)>1100$~GeV. Multijet backgrounds are reduced by requiring $\Delta 
\phi(jj)<1.0$ and events are removed if there are any jets with $p_T>30$~GeV 
between the two leading forward jets.
Figure~\ref{fig:CMS_vbf} shows the \etmiss~distribution after the full 
selection is applied.
Dominant backgrounds in this channel are $Z\to \nu\nu + \textrm{jets}$ and 
$W+\textrm{jets}$ whereas for the later the charged lepton is mis-reconstructed. 
These backgrounds are estimated from data control regions. Smaller background 
processes of single-top, top-pair and diboson production are estimated using 
simulations.

\begin{figure}[h!]
  \begin{minipage}{0.6\textwidth}
    \centering
    \includegraphics[scale=0.55]{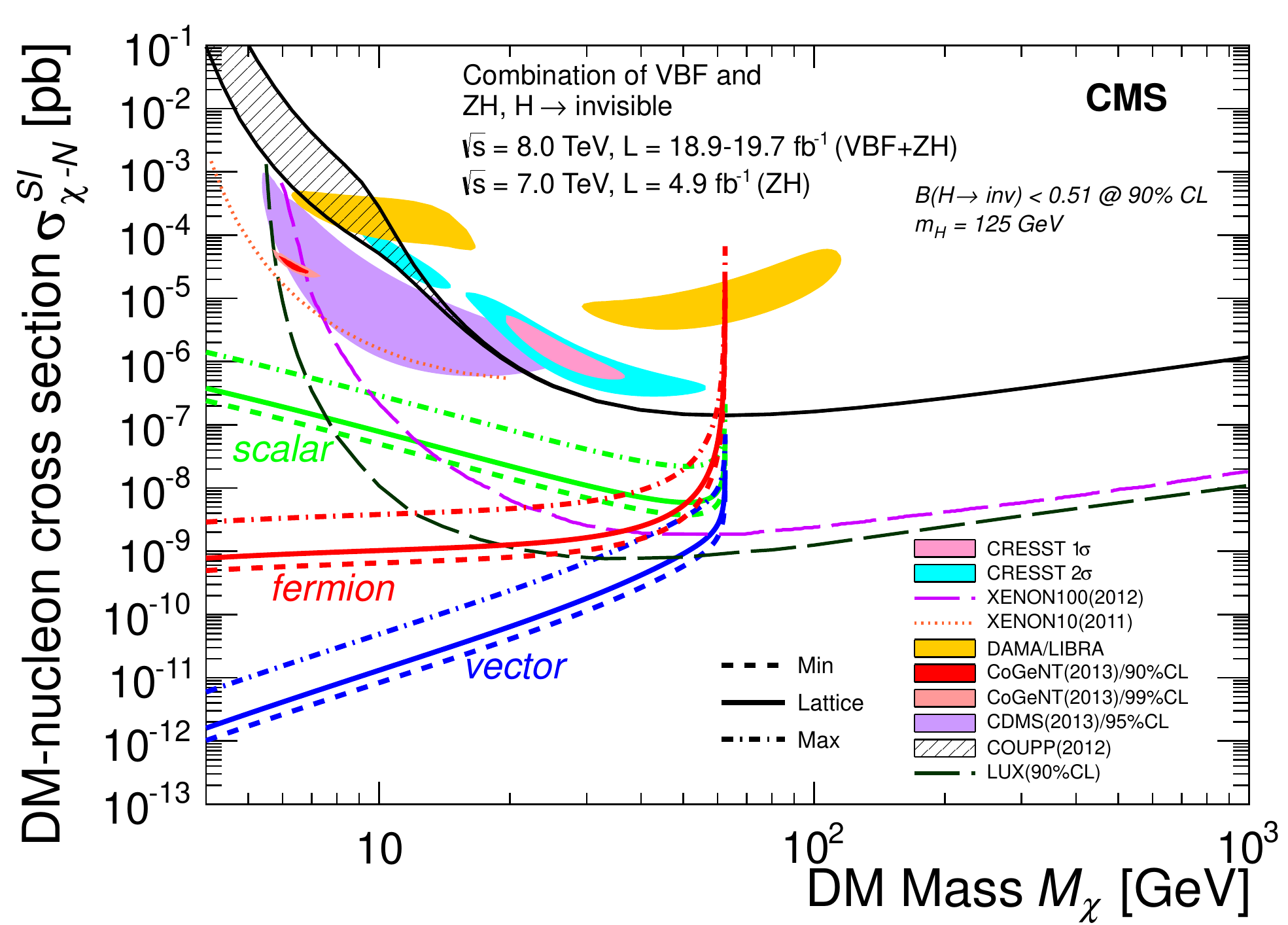}
  \end{minipage}
  \caption{Upper limits at 90$\%$ CL on the WIMP-nucleon cross section as a 
function of the DM mass, derived from the experimental upper limit on ${\cal 
BR}(H \to inv.)$ for $m_H = 125$~GeV, in the scenarios where the DM candidate is 
a scalar, a vector, or a Majorana fermion.~\cite{Chatrchyan:2014tja}}   
 \label{fig:CMS_higgs_limits}
\end{figure}

None of these searches yield any significant deviation from the background 
expectation and upper limits are set. These limits again are interpreted in the 
context of a Higgs-portal model as limit on the WIMP-nucleon scattering cross 
section for scalar, vector and fermionic dark matter candidates. 
Figure~\ref{fig:CMS_higgs_limits} shows the upper limits at 90$\%$ CL on the 
WIMP-nucleon cross section as a function of the DM mass, derived from the upper 
limit on ${\cal BR}(H \to inv.)$ for $m_H = 125$~GeV, in the scenarios where the 
DM candidate is a scalar, a vector, or a Majorana fermion.

\subsubsection{Limits derived from the total Higgs Branching Ratio}

The ATLAS collaboration \cite{ATLAS-CONF-2014-010} constrains invisible Higgs 
decays also using the combined measurements of $H\to\gamma\gamma,~ H \to ZZ \to 
4 \ell,~H \to WW \to \ell \nu\ \ell \nu, H\to \tau \tau, \textrm{and}~H\to 
b\bar{b} $ in combination with the upper limit on $ZH \to \ell \ell + \etmiss$ 
to derive an upper limit on the branching ratio to invisible states. Standard 
Model values for the couplings of the Higgs boson to massive particles are 
assumed for Higgs boson production and decay. Effective couplings to photons and 
gluons, $\kappa_\gamma,~\kappa_g$ respectively, are assumed to absorb potential 
loop contributions of new particles. The ratio of total width of the Higgs boson 
to Standard Model expectation is parametrized by $\kappa_H^2$:

\begin{eqnarray}
  \kappa^2_H & = & \Gamma_H/\Gamma_{H,\textrm{SM}} = \sum_i 
\kappa^2_i/(1-\mathcal{BR}_i)\\
  \sum_i \kappa_i^2 &=& 0.0023\kappa_\gamma^2+0.085\kappa_g^2+0.91
\end{eqnarray}

\begin{figure}[h!]           
  \begin{minipage}{0.6\textwidth}                                                                                                                                                 
   \centering                
   \includegraphics[scale=0.55]{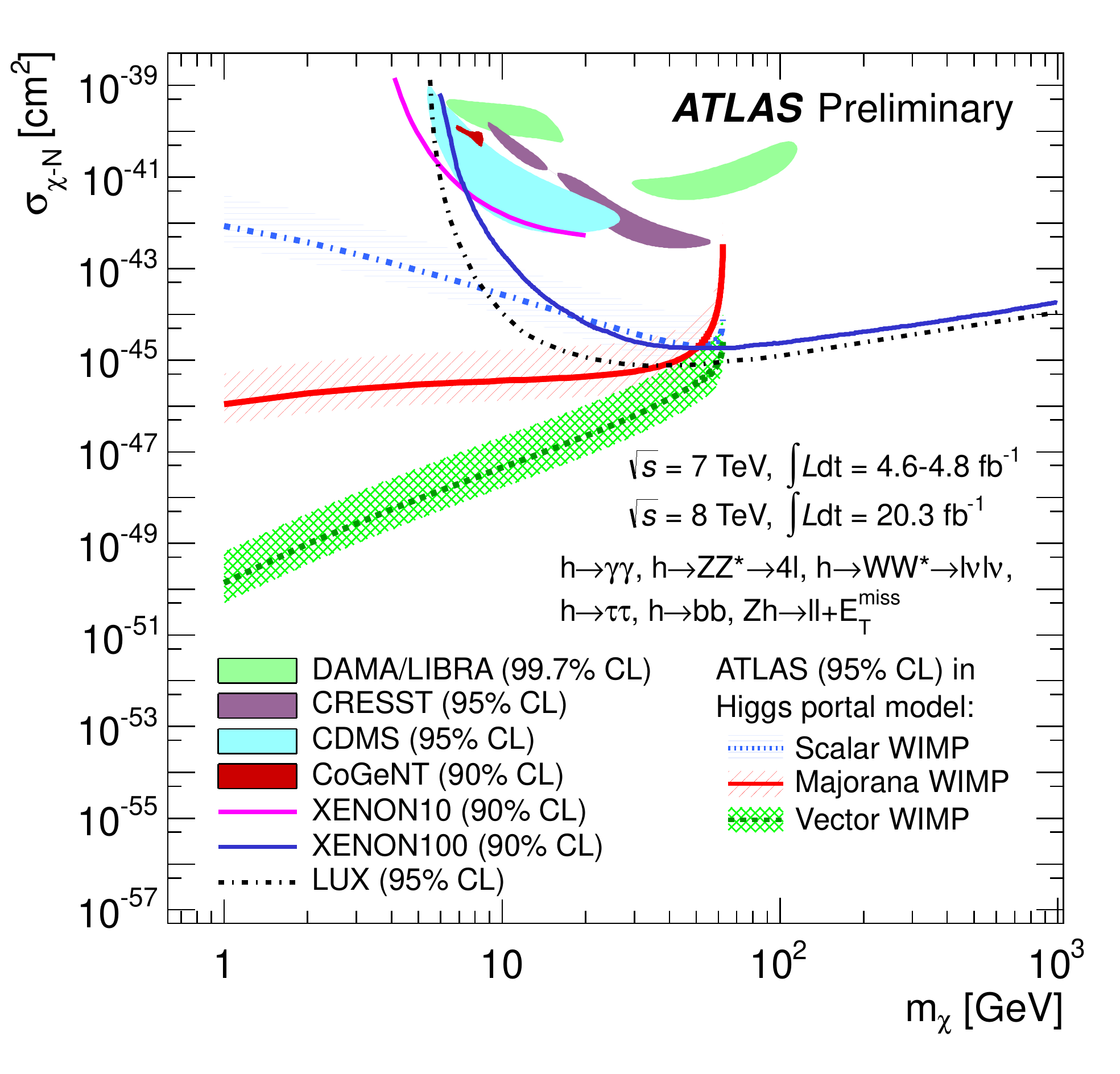}
  \end{minipage}             
  \caption{ ATLAS upper limit at 95$\%$ CL on the WIMP-nucleon scattering cross section in a Higgs portal model as a function of the mass of the dark matter particle, shown separately for a scalar, Majorana fermion, or vector boson WIMP. Excluded and allowed regions from direct detection experiments are also shown. These are spin-independent results obtained directly from searches for nuclei recoils from elastic scattering of WIMPs, rather than being inferred indirectly through Higgs boson exchange in the Higgs portal model.}
\label{fig:atlas_h_br2}    
\end{figure}

and a likelihood scan is performed. With constraints applied to ensure that the 
fit produces physical values the best fit yields $BR_i<0.37 ~(0.39)$ observed 
(expected) at $95\%$~CL using the combination of all channels.
Under the assumptions $2 \cdot m_\chi \leq m_H$ and the resulting Higgs boson 
decays to WIMP pairs account entirely for $BR_i$ limits are derived on the Higgs 
coupling to mass depending on the WIMP mass and parametrized as limit on the 
direct WIMP-nucleon scattering cross section via Higgs exchange. 
Figure~\ref{fig:atlas_h_br2} presents the limit on the spin-independent 
WIMP-nucleon scattering cross section.

\begin{figure}[h!]
  \begin{minipage}{0.9\textwidth}
    \centering
    \includegraphics[scale=0.65]{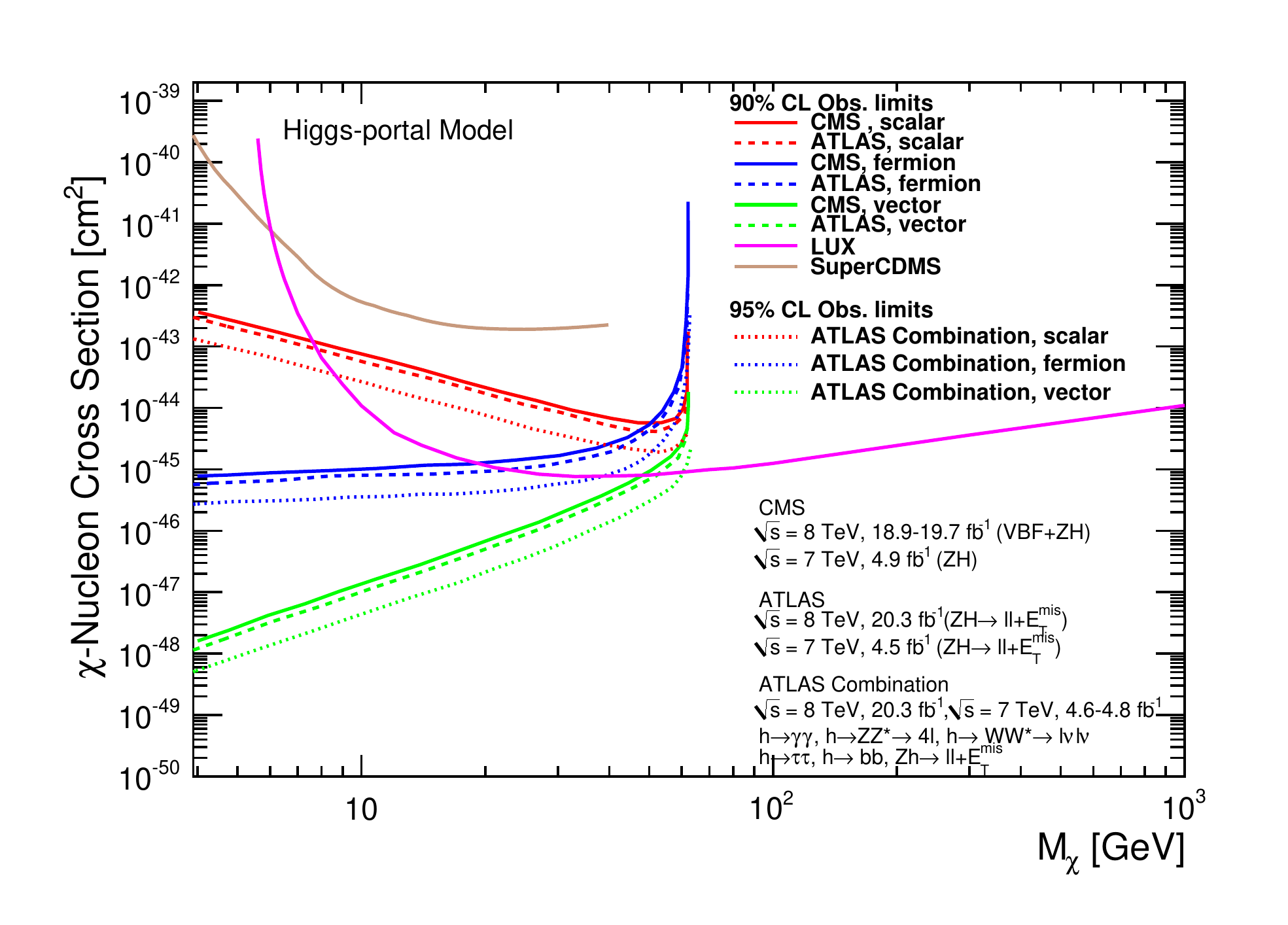}
  \end{minipage}
  \caption{ Observed 90$\%$ CL upper limit for WIMP-nucleon cross section with 
scalar, vector and Majorana DM particle from $Z(\to \ell \ell)H$ final state. These results 
from the CMS and the ATLAS collaboration are also compared with latest direct 
detection bound from LUX \cite{Akerib:2013tjd}, SuperCDMS \cite{Agnese:2014aze}.}
  \label{fig:zh_search}
\end{figure}

 
  The upper limits from ZH searches can be translated to WIMP-nucleon 
cross section. The current upper bound from these limits are similar for either 
collaboration, while the ATLAS limits perform slightly better. Comparison of 
these upper limits is shown in Figure ~\ref{fig:zh_search} for scalar, 
vector, Majorana fermion DM candidates. Although for $m_{\chi}> m_{H}/2$ the 
limits obtained by LUX are stronger compared to collider searches whereas at lower masses bounds from 
$ZH$ dominated for spin-independent interactions.
 
\section{Current Bounds}
\label{sec:results}
\begin{figure}[h!]
  \begin{minipage}{0.9\textwidth}
    \centering
    \includegraphics[scale=0.6]{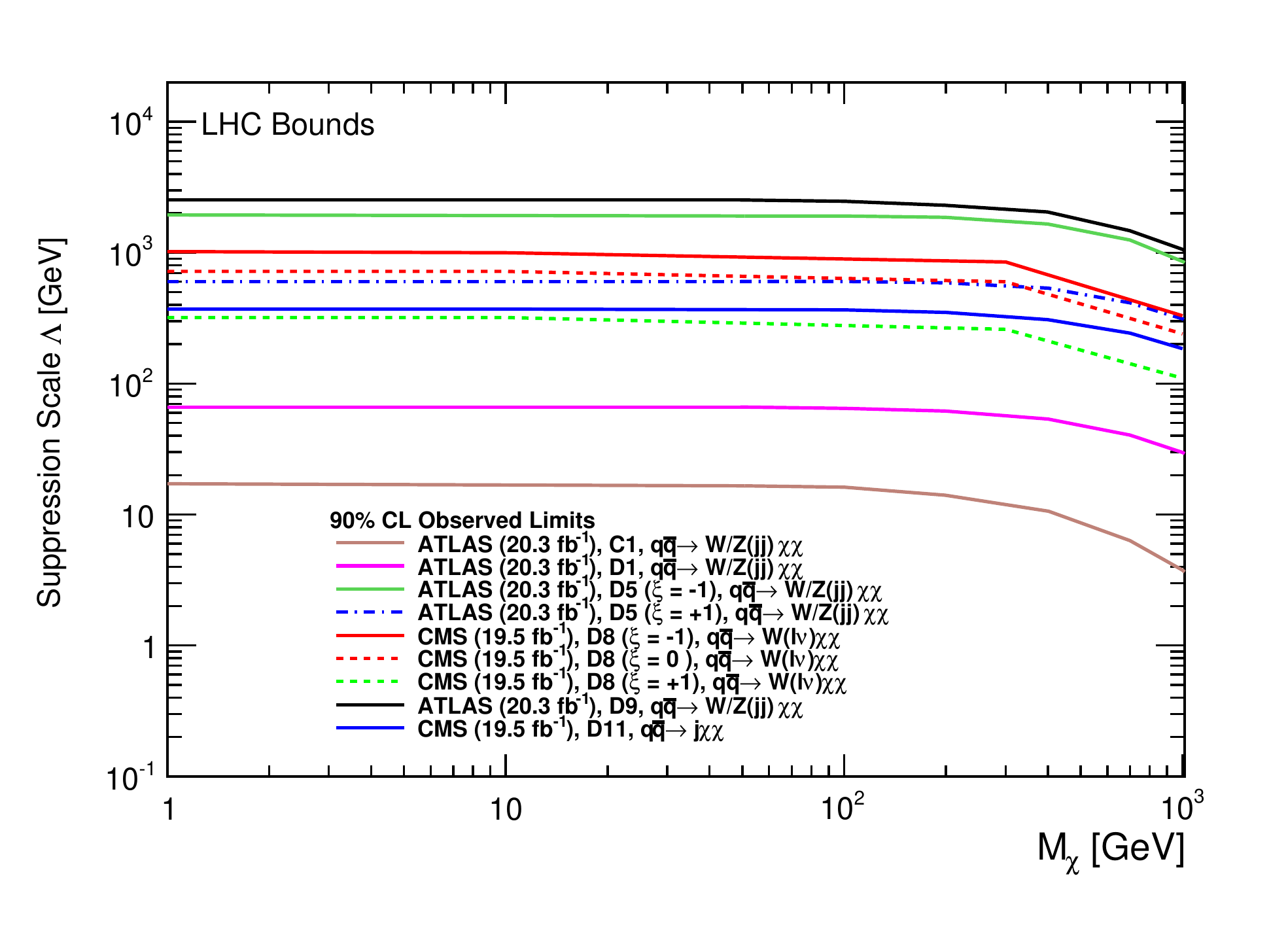}
  \end{minipage}
  \caption{The current bounds for different DM operators from LHC searches with different final state.}  
  \label{fig:bestLambda}
\end{figure}

  Figure~\ref{fig:bestLambda} shows the current 90$\%$CL lower  bounds on suppression scale $\Lambda$ as a function of  $M_{\chi}$ for various DM operators considered in different collider searches. These DM operators include C1, D1, D5, D8, D9 and D11. The ATLAS results from mono-$W/Z(jj)+\chi\chi$ gives the strongest lower limits for C1, D1, D5, and D9 DM operators, reflecting the impressive control of systematics possible in boosted boson searches. The best bounds for the D11 operator comes from the mono-jet search done by the CMS collaboration (as expected for an operator coupling only to gluons), while strongest bound for D8 are observed in mono-$W(\ell \nu)+\chi\chi$, as it was not included in the hadronic decay search of ATLAS. For D5 and D8 DM operators results are presented for different coupling scenarios that determine the type of interference as discussed in Sec.\ref{sec:exp}. 
  
\begin{figure}[h!]
  \begin{minipage}{0.9\textwidth}
      \centering 
      \includegraphics[scale=0.6]{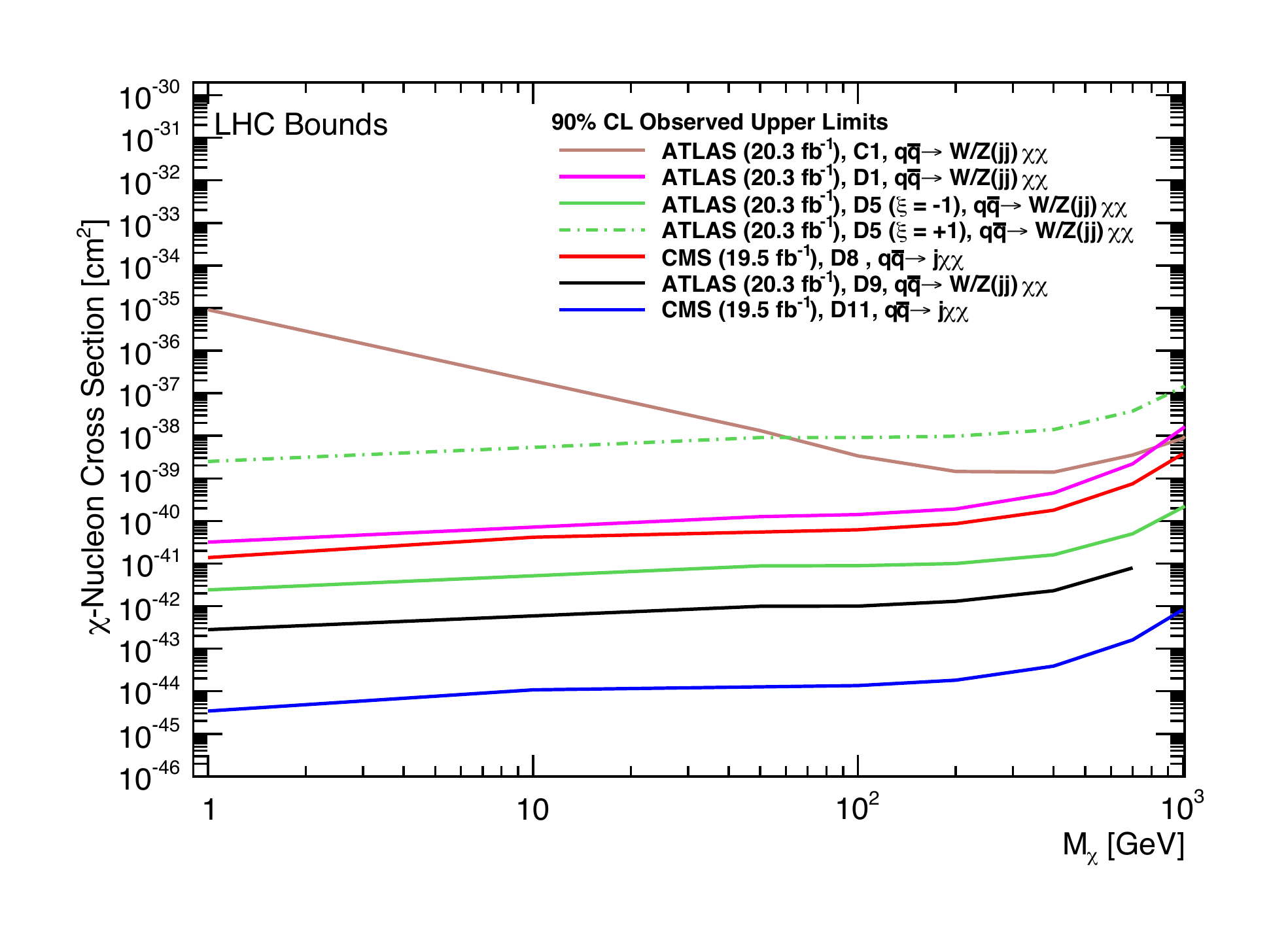}
  \end{minipage}
  \caption{The current bounds on upper limit for $\chi$-nucleon cross section for different DM operators from collider searches with different final state.} 
  \label{fig:best_xs}
\end{figure}

  The current status of WIMP-nucleon cross section upper limit from collider searches is presented in Figure~\ref{fig:best_xs} for different DM operators. The best bounds for each type of operator are from the similar final state as in case of $\Lambda$. For D8 operator the best upper limits are from the CMS collaboration in mono-jet final state. 


The different searches summarized here have not found any evidence of DM production. The WIMP-nucleon cross section limits from collider are comparable to direct searches. Present lower bounds on the scale ($\Lambda$) of these interactions vary from few GeV (for C1) to few TeVs (for D9) for different type of DM operators and couplings discussed in this article. 


\section*{Acknowledgements}

The research of WS is partly supported by the US Department of Energy under 
contract DE-FG02-04ER41268.  SC and MT are supported by the grant DE-SC000999.
AA is supported by the US Department of Energy under contract DE-FG02-13ER41942.

\bibliography{bibliography}

\end{document}